\documentclass[useAMS,usenatbib]{mnras}
\usepackage{amssymb,amsmath,epsfig,times,natbib,longtable,color,graphicx,dcolumn,txfonts}
\long\def\symbolfootnote[#1]#2{\begingroup%
\def\thefootnote{\fnsymbol{footnote}}\footnote[#1]{#2}\endgroup} 
\def\aj{AJ}
\def\araa{ARA\&A}
\def\apj{ApJ}
\def\apjl{ApJ}
\def\apjs{ApJS}
\def\aap{A\&A}
\def\aaps{A\&AS}
\def\mnras{MNRAS}
\def\pasp{PASP}
\def\pasj{PASJ}
\def\nat{Nature}
\def\iaucirc{IAU~Circ.}

\newcommand{\nustar}{\textit{NuSTAR}}

\newcommand{\swift}{{\it Swift}}
\newcommand{\xmm}{{\it XMM-Newton}}
\newcommand{\rosat}{{\it ROSAT}}

\makeatletter
 \def\hlinewd#1{%
   \noalign{\ifnum0=`}\fi\hrule \@height #1 \futurelet
    \reserved@a\@xhline}
\makeatother

\newcommand{\kms}{km\,s$^{-1}$}

\title[The changing look AGN HE~1136-2304]{The detection and X-ray view of the changing look AGN HE~1136-2304}

\author[M. L. Parker et al.]{M. L. Parker$^{1}$,\thanks{Email: mlparker@ast.cam.ac.uk}
  S. Komossa,$^{2}$
  W. Kollatschny,$^{3}$
  D. J. Walton,$^{4,5}$
  N. Schartel,$^{6}$\newauthor
  M. Santos-Lle\'o,$^{6}$
  F. A. Harrison,$^{5}$
  A. C. Fabian,$^{1}$
  M. Zetzl,$^{3}$
  D. Grupe$^{7}$\newauthor
  P. M. Rodr\'iguez-Pascual$^{6}$
  and R. V. Vasudevan$^{1}$\\  
  $^{1}$Institute of Astronomy, Madingley Road, Cambridge, CB3 0HA\\
  $^{2}$Max-Planck-Institut f\"ur Radioastronomie, Auf dem H\"ugel 69, D-53121 Bonn, Germany\\
  $^{3}$Institut f\"ur Astrophysik, Universit\"at G\"ottingen, Friedrich-Hund-Platz 1, D-37077 G\"ottingen, Germany\\
  $^{4}$Jet Propulsion Laboratory, California Institute of Technology, 4800 Oak Grove Drive, Pasadena, CA 91109, USA\\
  $^{5}$California Institute of Technology, 1200 East California Boulevard, Pasadena, CA 91125, USA\\
  $^{6}$XMM-Newton SOC, ESA, ESAC, Villafranca del Castillo, Camino Bajo del Castillo s/n, E-28692 Villanueva de la Ca{\~n}ada Madrid, Spain\\
  $^{7}$Department of Earth and Space Science, Morehead State University, 235 Martindale Drive, Morehead, KY 40351, USA\\
}
\date{}

\begin{document}

\maketitle

\begin{abstract}
We report the detection of high-amplitude X-ray flaring of the AGN HE~1136-2304, which is accompanied by a strong increase in the flux of the broad Balmer lines, changing its Seyfert type from almost type 2 in 1993 down to 1.5 in 2014. HE~1136-2304 was detected by the \xmm\ slew survey at $>10$ times the flux it had in the \rosat\ all-sky survey, and confirmed with \swift\ follow-up after increasing in X-ray flux by a factor of $\sim30$. Optical spectroscopy with SALT shows that the AGN has changed from a Seyfert 1.95 to a Seyfert 1.5 galaxy, with greatly increased broad line emission and an increase in blue continuum AGN flux by a factor of $>4$. The X-ray spectra from \xmm\ and \nustar\ reveal moderate intrinsic absorption and a high energy cutoff at $\sim100$~keV. We consider several different physical scenarios for a flare, such as changes in obscuring material, tidal disruption events, and an increase in the accretion rate. We find that the most likely cause of the increased flux is an increase in the accretion rate, although it could also be due to a change in obscuration.
\end{abstract}


\section{Introduction}

Outbursts in radio-quiet active galactic nuclei (AGN) are a relatively unexplored area of AGN physics. However, observations of these unusual events have the potential to trace accretion physics in detail, particularly when multiple instruments can be used to observe at different wavelengths. 
In recent years, there have been several detections of AGN in anomalously low or high X-ray states using \swift\ monitoring or  \xmm\ 
surveys \citep[e.g.][]{Schartel07, Schartel10, Grupe12, Grupe13_wpvs007, Miniutti09_phl1092, Miniutti13, Saxton14, Parker14_mrk1048, 
Parker14_mrk335, Gallo14, Grupe15, Komossa15_seyferts}. Despite the high-amplitude X-ray variability (factors $>$ 20--100) of AGN, these events rarely come with strong changes in the optical emission lines, implying that the observed changes mostly happen along our line-of-sight, and do not affect the bulk of the emission-line regions. An interesting example is GSN~069, which was detected during an \xmm\ slew observation in 2010 at a soft X-ray flux level at least 240 times that which it had when \rosat\ failed to detect it in 1994 \citep{Miniutti13}. The X-ray spectrum is extremely soft and unobscured, and it shows rapid soft X-ray variability, which is consistent with Seyfert 1 behaviour. However, it has no detected broad line emission, so can be classified as a Seyfert 2.

The term `changing look' was coined to refer to X-ray observations of Compton-thick AGN becoming Compton-thin, and vice-versa \citep{Guainazzi02_ngc6300, Guainazzi02_ugc4203, Matt03, Marchese12, Miniutti14, Ricci16}. This transition can be extremely rapid: \citet{Risaliti05} found NGC~1365 switching from Compton thick to Compton thin in under 6 weeks.
There are also well documented cases of changing-look behaviour in the optical band, where the optical spectra of AGN change so as to move to a different Seyfert classification. Cases include NGC~3515 \citep{Collin-Souffrin73}, NGC~7306 \citep{Tohline76}, NGC 4151 \citep{Penston84}, Fairall 9 \citep{Kollatschny85}, Mrk 1018 \citep{Cohen86}, Mrk 99 \citep{Tran92}, NGC 1097 \citep{Storchi-Bergmann93}, NGC 7582 \citep{Aretxaga99}, NGC 2617 \citep{Shappee14}, Mrk 590 \citep{Denney14}, and the quasars SDSS\,J015957.64+003310.5 \citep{LaMassa15} and SDSS\,J101152.98+544206.4 \citep{Runnoe16}.
The most extreme examples known come with dramatic changes not only in their broad Balmer lines, but also in their more narrow high-ionization lines like [FeVII] and [FeXIV].  Cases include IC 3599 \citep{Brandt95, Grupe95, Komossa99}, SDSS\,J095209.56+214313.3 \citep{Komossa08} and SDSS\,J074820.67+471214.3 \citep{Wang11}. More recent studies have begun to look at small samples of changing look AGN \citep{Ruan15,MacLeod16, Runco16}.

There are several potential causes for rapid changes in the (apparent) brightness of AGN. 
In the unified model of AGN, the differences between Seyfert 1 and Seyfert 2 galaxies are interpreted as being largely due to the effect of inclination, with the inclination of Seyfert 2 AGN being such
 that the torus obscures the inner regions, strongly absorbing the X-ray emission and blocking 
the broad line region (BLR) from the observer. When the observer's line of sight skims the edge of the torus, rapid changes in the observed flux can occur when obscuring clouds at the edge of the torus move to allow a clear view of the central regions of the AGN \citep[e.g.][]{Goodrich89,Antonucci93,Leighly15}. However, it is challenging to attenuate a large fraction of the light from the BLR with this mechanism, as this requires different parts of the extended torus to act simultaneously.  Alternatively, these events could be caused by large changes in the accretion rate of the AGN, with a large increase in the accretion rate correspondingly increasing the AGN luminosity \citep[e.g.][]{Nicastro00,Korista04,Elitzur14, Runnoe16}. Finally, stellar tidal disruption events \citep[TDEs;][]{Rees88,Komossa15_TDEs} could cause a sudden increase in brightness. Systematic searches for changing look AGN in quasars suggest that the majority are related to changes in the accretion rate \citep{Ruan15} and $>15$\% of strongly-variable luminous quasars show such variability on time-scales of 8--10~years \citep{MacLeod16}, although the sample sizes in these studies are relatively small.

In this work we present \xmm , \nustar , and SALT observations of the flaring AGN HE~1136-2304, 
with the aim of establishing the cause of the outburst and its change in Seyfert type. 
HE~1136-2304 is a nearby, relatively unknown AGN at redshift $z$=0.027 \citep[][RA: 11h38m51.1s DEC: -23d21m36s]{Reimers96}. It was detected in X-rays in the ROSAT All-Sky Survey \citep[RASS;][]{Voges00} and is a faint radio source \citep[8.2 mJy;][]{Condon98}.

Throughout this paper, we assume a cosmology with $H_0=70$~km~s$^{-1}$~Mpc$^{-1}$, $\Omega_\Lambda=0.73$ and $\Omega_\textrm{M}=0.27$.

\section{Observations and Data Reduction}
\label{section_datareduction}


\subsection{X-ray observations}
\label{section_xray_observations}

The X-ray observations presented here are based off \xmm\ \citep{Jansen01} proposal ID 74126 (PI N. Schartel), \emph{`Outbursts of Radio-Quiet AGN'}. This proposal was intended to obtain an observation of a flaring AGN, and as such had a triggering condition that the source flux had to be in excess of 15 times the flux observed with \rosat . HE~1136-2304 was initially detected in outburst using the \xmm\ Slew Survey in 2010, at a flux ratio of $13.3\pm2.0$. While significantly above the \rosat\ level, this did not meet the required flux level. However, follow up observations with \swift\ found that the source reached a flux level of $\sim16$ times that found by \rosat , representing a genuine outburst, and a peak flux ratio of $\sim30$ was later found with \swift. Based on this detection, we triggered the simultaneous observations with \xmm\ and \nustar\ \citep{Harrison13}. The flux ratios for all observations are given in Table~\ref{table_ratios}. We note that the source was also detected in the \swift\ BAT 70-month catalogue, with an average 14--195~keV flux of $17_{-11}^{+22}\times 10^{-12}$~erg~s$^{-1}$~cm$^{-2}$.

\begin{table*}
\centering
\caption{\rosat\ PSPC, \xmm\ EPIC-pn and \swift\ XRT fluxes for HE\,1136-2304.}
\label{table_ratios}%
\begin{tabular}{l c c c c}
\hline
\hline
Instrument & Date & Count rate$^*$ &0.2--2~keV Flux & Ratio\\
\hline
\rosat\ & 1990 &  $0.042\pm0.015$ & $0.4\pm0.2$ & 1\\
\xmm\ (slew)  & 2007-01-12 & $<1.5$ & $<2.1$ & $<5.3$ \\
\xmm\ (slew) & 2010-01-03 & $3.7\pm0.6$ & $5.3\pm0.8$ & $13.3\pm2.0$ \\
\xmm\ (slew)  & 2011-12-10 & $2.0\pm0.5$ & $2.9\pm0.7$ & $7.3\pm1.7$ \\
\xmm\ (slew)  & 2014-06-06 & $3.3\pm0.9$ & $4.7\pm1.2$ & $11.8\pm3.1$ \\
\swift $_\textrm{initial}$ (PC) & 2014-06-25 & $0.46\pm0.02$ &$6.2\pm0.3$ & $15.5\pm1.3$\\
\xmm\ & 2014-07-02 & $2.67\pm0.01$ & $4.32\pm0.01$ & $10.8\pm0.03$ \\
\swift $_\textrm{max}$ (WT) & 2014-08-01 & $0.93\pm0.04$ & $11.6\pm0.5$ & $29\pm1$\\

\hline
\hline
\end{tabular}

Fluxes are in units of $10^{-12}$~erg~s$^{-1}$~cm$^{-2}$, and are calculated using model 2 from Table~\ref{table_xrayfit}, including the effect of absorption. Ratios are relative to the ROSAT flux.
$^*$Energy bands for the count rates differ between instruments. For \xmm , all rates are from 0.2--2~keV. For \rosat , the count rate is from 0.1--2.4~keV, and for \swift , the count rate is from 0.3--10~keV.
\end{table*}

We used the \xmm\ Science Analysis Software (SAS) version 14.0.0 to reduce the \xmm\ data, extracting EPIC-pn \citep{Struder01} and MOS \citep{Turner01_MOS} event files using \textsc{epproc} and \textsc{emproc} respectively. Both the EPIC-pn and MOS detectors were operated in small window mode. We filter for background flares, and extract source and background spectra from $30^{\prime\prime}$ circular extraction regions, choosing the background region so as to avoid contaminating photons from HE~1136-2304 or other objects in the field of view. We use the \textsc{specgroup} tool to rebin the spectra to a signal-to-noise ratio of 6 and oversampling the instrumental resolution by a factor of 3, ensuring the applicability of $\chi^2$ statistics.
A preliminary investigation of the reflection grating spectrometer (RGS) spectra reveals no highly significant features, so we restrict our analysis to the lower resolution, higher sensitivity detectors.

\begin{figure}
\centering
\includegraphics[width=\linewidth]{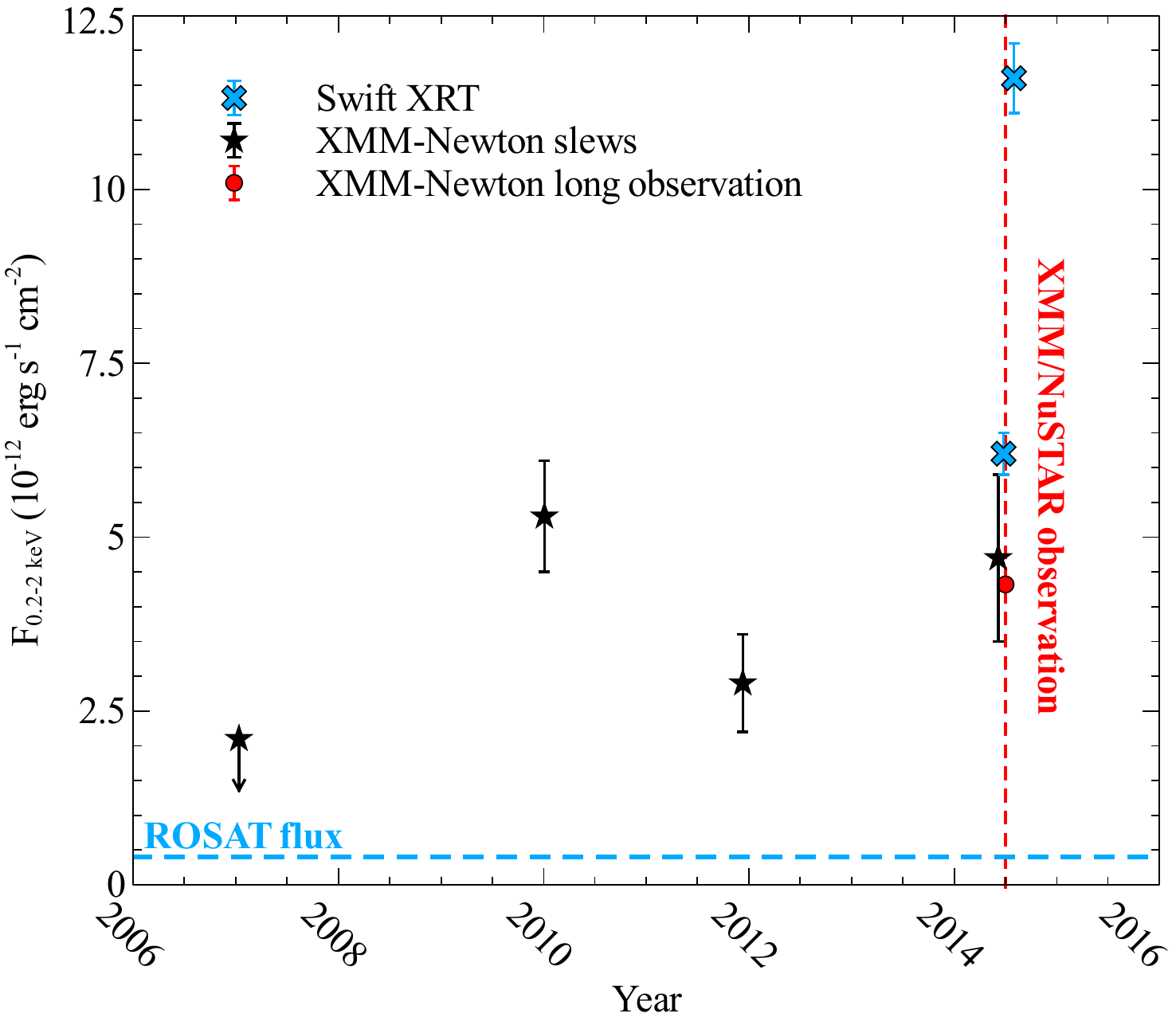}
\caption{Long-term lightcurve showing the fluxes from Table~\ref{table_obsids}. The 1990 ROSAT flux is shown by the horizontal dashed line and the time of the joint \xmm /\nustar\ observation is shown by the vertical dashed line. The SALT spectrum was taken 3 days after the end of the \nustar\ observation.}
\label{fig_longterm}
\end{figure}

The \nustar\ data were reduced using the \nustar\ Data Analysis Software (\textsc{nustardas}) version 1.4.1 and the CALDB version 20150316 (more recent versions at the time of writing only update the \nustar\ clock correction file, so will not affect our results). Spectra and lightcurves were extracted with the \textsc{nuproducts} tool, using 60$^{\prime\prime}$ circular extraction regions. As with the \xmm\ spectra, the \nustar\ spectra are binned to oversample the data by a factor of 3 and a signal-to-noise ratio of 6. We fit the \nustar\ spectra over the whole band (3--79~keV). The details of these observations, and those of \xmm, are given in Table~\ref{table_obsids}, and the fluxes are plotted in Fig.~\ref{fig_longterm}.

\begin{table*}
\centering
\caption{Observation details for the \xmm\ and \nustar\ detectors}
\label{table_obsids}
\begin{tabular}{l l l l c l}
\hline
Instrument& Obs ID & Start time (UTC) & End time (UTC) & Clean exposure time (ks) &Count rate (s$^{-1}$)\\
\hline
EPIC-pn & 0741260101 & 2014-07-02 08h08m & 2014-07-03 13h56m & 71.11 & $3.743\pm0.007$\\
EPIC-MOS1&0741260101 & 2014-07-02 08h02m & 2014-07-03 13h52m & 97.20 & $0.876\pm0.003$\\
EPIC-MOS2&0741260101 & 2014-07-02 08h03m & 2014-07-03 13h52m & 97.02 & $1.165\pm0.003$\\
FPMA &80002031002    & 2014-07-02 08h16m & 2014-07-02 22h31m & 23.81 & $0.308\pm0.004$\\
     &80002031003    & 2014-07-02 22h31m & 2014-07-04 10h01m & 63.57 & $0.255\pm0.002$\\
FPMB &80002031002    & 2014-07-02 08h16m & 2014-07-02 22h31m & 23.80 & $0.292\pm0.004$\\
     &80002031003    & 2014-07-02 22h31m & 2014-07-04 10h01m & 63.52 & $0.245\pm0.002$\\
\hline
\end{tabular}
\end{table*}

All X-ray spectra are fit using \textsc{Xspec} version 12.8.2l, and all errors are reported at 1$\sigma$ unless otherwise stated. We assume Galactic absorption of $3.82\times10^{20}$~cm$^{-2}$ \citep{Willingale13}, and use the abundances of \citet{Wilms00}.

\subsection{Optical spectroscopy with the SALT telescope}
To determine if the X-ray flare corresponded to a changing look in the optical band, we observed HE\,1136-2304 with the 10m Southern
African Large Telescope (SALT)
nearly simultaneously to the XMM observations on July 7, 2014
under photometric conditions.
The optical spectrum was taken with the Robert Stobie Spectrograph
 attached to the telescope
 using the PG0900
grating with a 2.0$^{\prime\prime}$ wide slit. 
We covered the wavelength range from
4203 to 7261~\AA\  at a spectral resolution of 6.5~\AA\ (FWHM) and a reciprocal
dispersion of 0.98~\AA\ pixel$^{-1}$. The observed wavelength range corresponds
to a wavelength range from 4078--7050~\AA\ in the rest frame of the galaxy. 
 There are two gaps in the spectrum caused by the gaps between the three CCDs:
one between the blue and the central CCD chip as well as one between the
central and red CCD chip covering the wavelength ranges
5206.5--5262.7~\AA\  and 6254.4--6309.1~\AA\ (5079--5135~\AA\ and 6100--6150~\AA\
in the rest frame).
The seeing was 1.1$^{\prime\prime}$, and the exposure time was $2\times600$ seconds (20 minutes).

In addition to the galaxy spectrum, necessary flat-field and
Xe arc frames were also observed, as well
as a spectrophotometric standard star for flux calibration (LTT4364).
The spectrophotometric standard star
was used to correct the measured counts for the combined
transmission of the instrument, telescope and atmosphere
as a function of wavelength.
Flat-field frames
were used to correct for differences in sensitivity both
between detector pixels and across the field.
The spatial
resolution per binned pixel is 0\arcsec\hspace*{-1ex}.\hspace*{0.3ex}2534
for our SALT spectrum.
We extracted 7 columns from our object spectrum 
corresponding to 1\arcsec\hspace*{-1ex}.\hspace*{0.3ex}77.

The reduction of the spectra (bias subtraction, cosmic ray correction,
flat-field correction, 2D-wavelength calibration, night sky subtraction, and
flux calibration) was done in a homogeneous way with IRAF reduction
packages \citep[e.g.][]{kollatschny01}. 
We corrected the optical spectra of HE1136-2304
for Galactic extinction.
We used the reddening value E(B-V) = 0.03666 deduced from
 the \citet{schlafly11} re-calibration of the 
\citet{schlegel98} infrared-based dust map. The
reddening law of \citet{fitzpatrick99} with R$_{V}$\,=\,3.1
was applied to our spectra.
All wavelengths were converted to the rest frame of the galaxy (z=0.027). 

\section{Results}
\subsection{X-ray spectrum}
\label{section_xray_results}
We run a series of tests on the X-ray spectrum of HE~1136-2304, using the excellent broad-band data to try and shed light on the nature of the changes in the optical spectrum. We look for changes in the flux and spectral shape within the observations, which may indicate what dominates the variability on short time-scales, we search for the signatures of absorption (either ionized or neutral) in the spectrum, and fit the spectrum with various models that can be used to compare it to other, similar, AGN.

\begin{figure*}
\centering
\includegraphics[height=5cm]{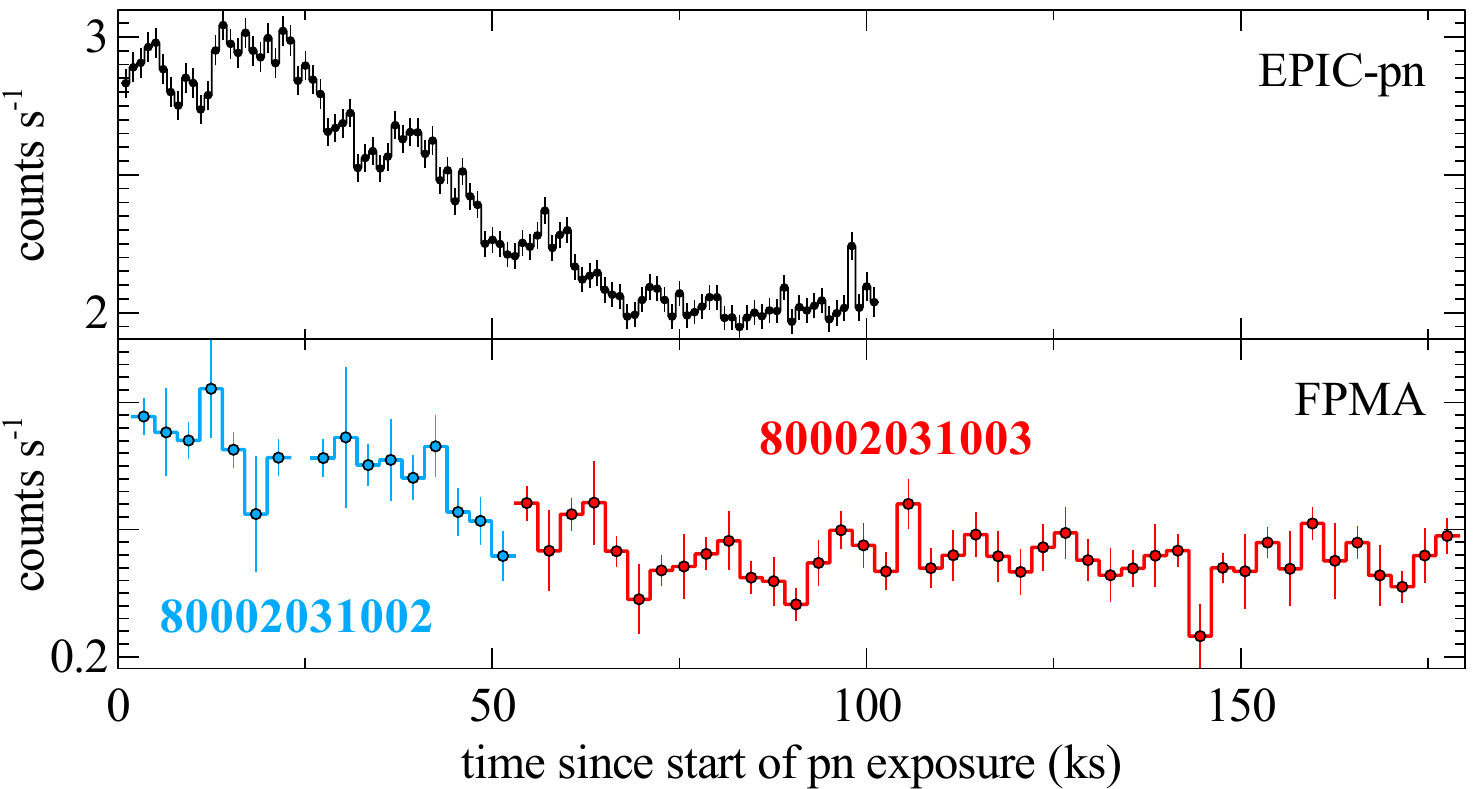}
\includegraphics[height=5cm]{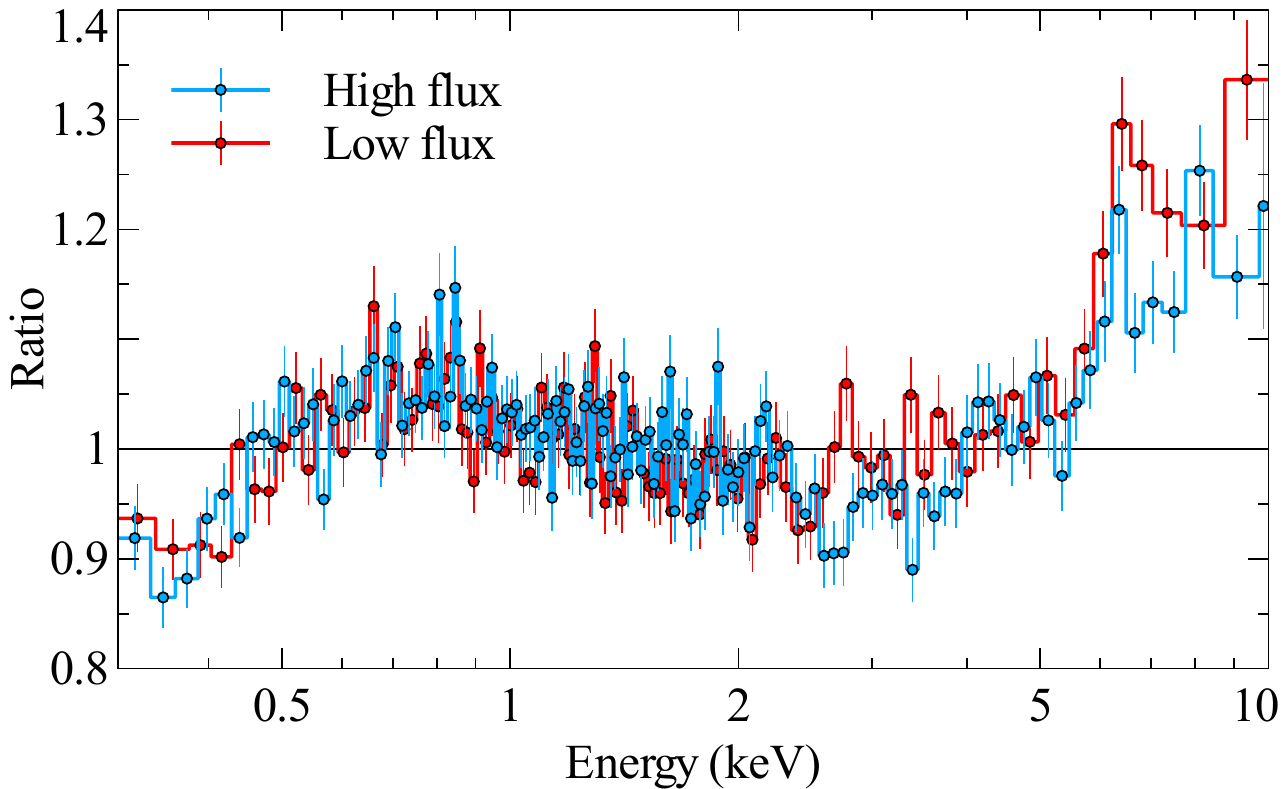}
\caption{Left: Background subtracted \xmm\ EPIC-pn and \nustar\ FPMA light curves over the whole energy range, binned to 1 and 3~ks, respectively. The two \nustar\ observation IDs are plotted separately, in red and blue. Right: Ratio of the EPIC-pn data to an absorbed power law for the two halves (high and low flux) of the \xmm\ observation, showing the absence of spectral changes with flux.}
\label{fig_lightcurves_comparison}
\end{figure*}

In Fig.~\ref{fig_lightcurves_comparison} we show the \xmm\ and \nustar\ light curves over the whole observation. A $\sim30$ per cent drop in flux is immediately obvious in the \xmm\ EPIC-pn light curve, which declines smoothly over around 25--75~ks. A similar, though less strong, drop is also observed in the \nustar\ lightcurve, which declines by around 25 per cent. The high and low flux intervals are (coincidentally) neatly divided by the two \nustar\ observation IDs, so the high energy spectral evolution, if present, should be visible in the difference between the two spectra.

While there is very little short-term variability in the lightcurve (Fig.~\ref{fig_lightcurves_comparison}, left), there is a $\sim30$~\% drop in flux in the 0.3--10~keV band over the course of the \xmm\ observation. In the right panel of Fig.~\ref{fig_lightcurves_comparison} we show the ratio of the EPIC-pn spectrum to an absorbed power law for the two halves of the \xmm\ observation, allowing for a difference in normalization between the two. It is apparent from this that there is no significant change in spectral shape corresponding to the drop in flux. The only potential difference is a slight increase in the iron line residuals, as would be expected if the narrow line originates far from the X-ray continuum source and therefore responds slowly. Because of the lack of spectral evolution, we combine the two \nustar\ observations into a single spectrum for each of the two FPMs, and fit all the data simultaneously.

\begin{figure}
\centering
\includegraphics[width=\linewidth]{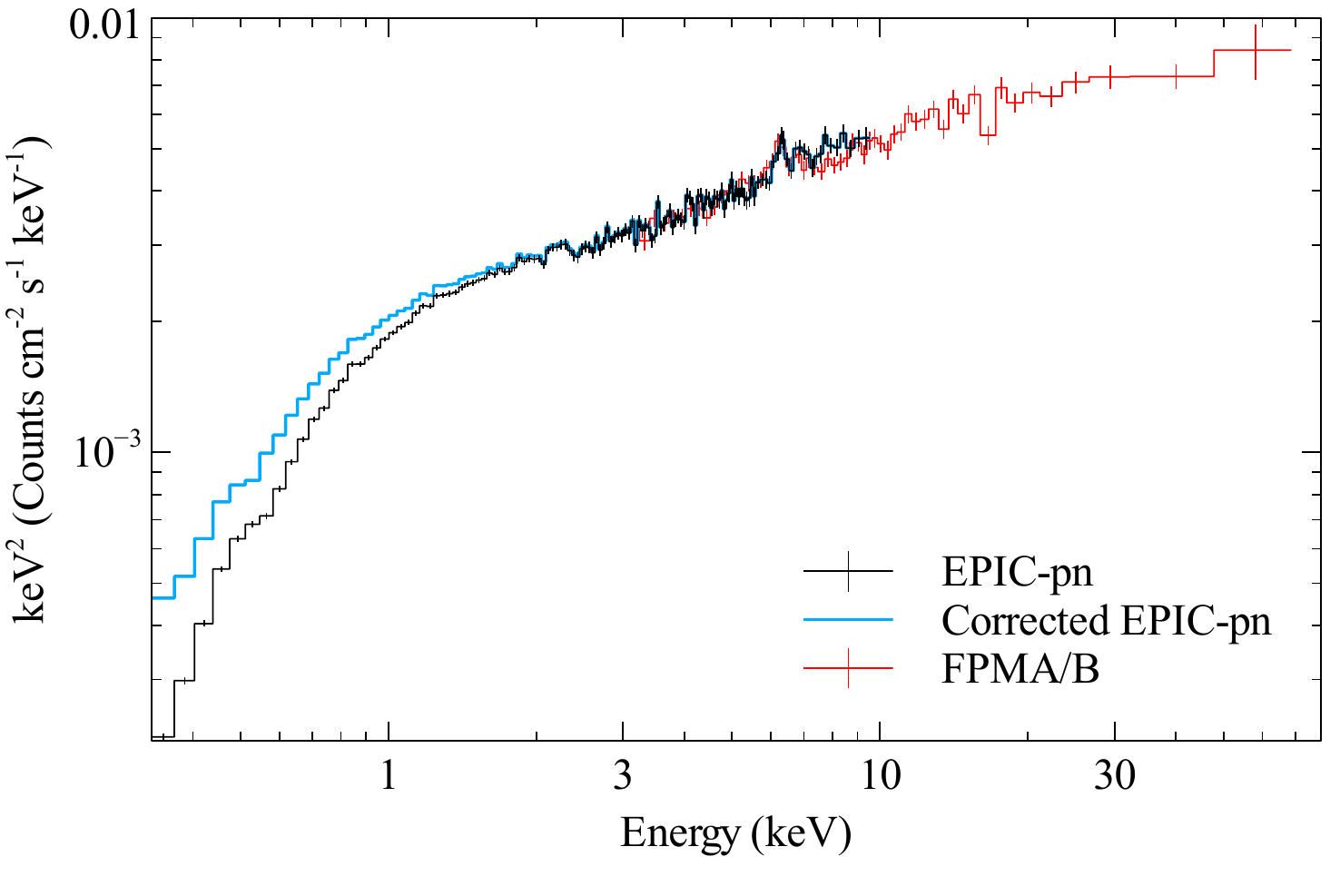}
\caption{Unfolded \xmm\ and \nustar\ spectrum of HE\,1136-2304 to a power law with $\Gamma=0$. The blue line shows the EPIC-pn points after correcting for Galactic absorption. It is obvious from this that there is additional absorption intrinsic to the source.}
\label{fig_broadband_unfolded}
\end{figure}

\begin{figure}
\centering
\includegraphics[width=\linewidth]{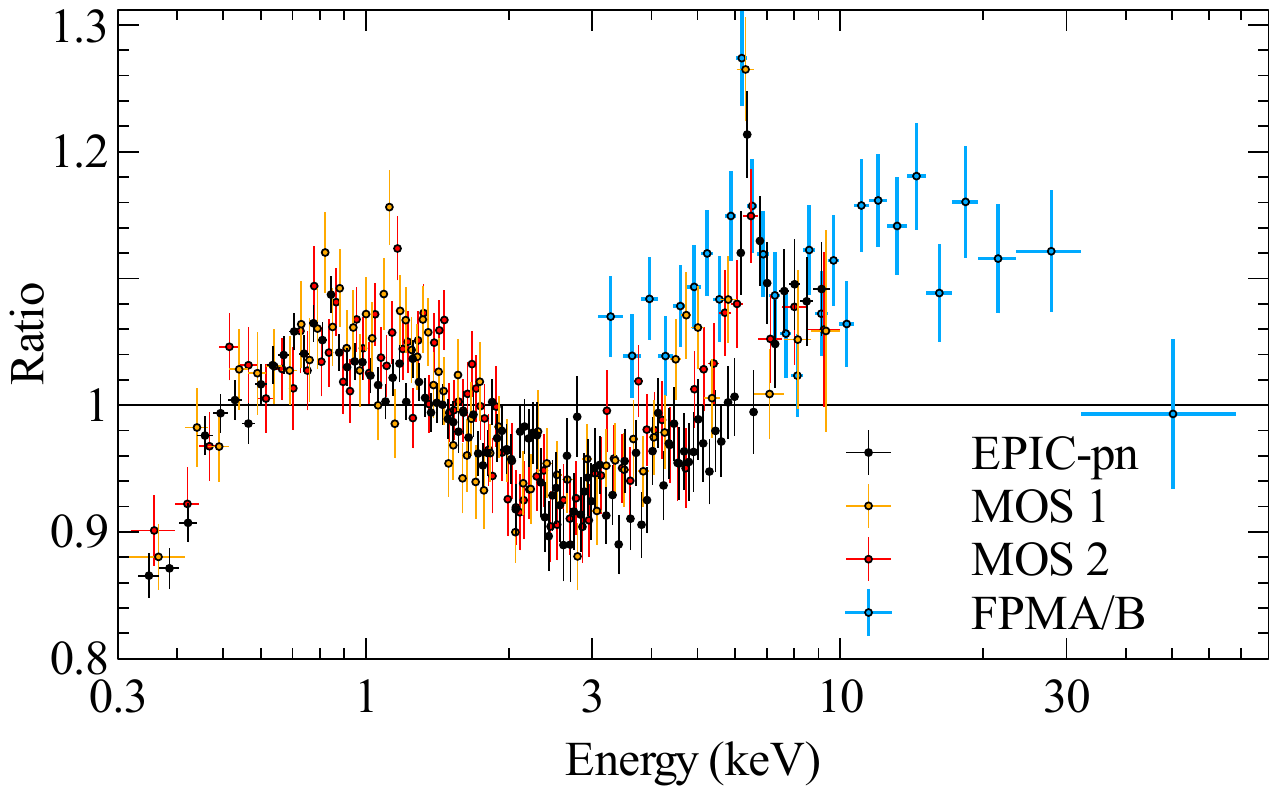}
\caption{Ratio of the X-ray spectrum to an absorbed power law, fit from 0.3--80~keV. The soft excess, iron line and hard excess/cutoff are all clearly visible. \nustar\ FPMA and FPMB data are grouped in \textsc{Xspec} and all spectra are rebinned for clarity.}
\label{fig_broadband_ratio}
\end{figure}

In Fig.~\ref{fig_broadband_unfolded} we show the EPIC-pn and focal plane module (FPM) A/B spectra, unfolded to a $\Gamma=0$ power law. The spectrum is fairly hard, and shows significant line emission around 6.4~keV. We also show the pn spectrum after correcting for Galactic absorption. A decrease in flux at low energies is still apparent, implying that there is a significant absorbing column intrinsic to the source.
We next examine the residuals after taking into account this additional absorption. In Fig.~\ref{fig_broadband_ratio} we show the ratio of the broad-band X-ray spectrum to the same absorbed power law, including both Galactic and intrinsic absorption (\emph{tbabs$\times$ztbabs$\times$powerlaw} as an \textsc{Xspec} model, i.e. Galactic absorption$\times$intrinsic absorption$\times$power law). A strong narrow line is clearly visible at 6.4~keV, along with a soft excess and a turnover in the \nustar\ band.

\begin{figure}
\centering
\includegraphics[width=\linewidth]{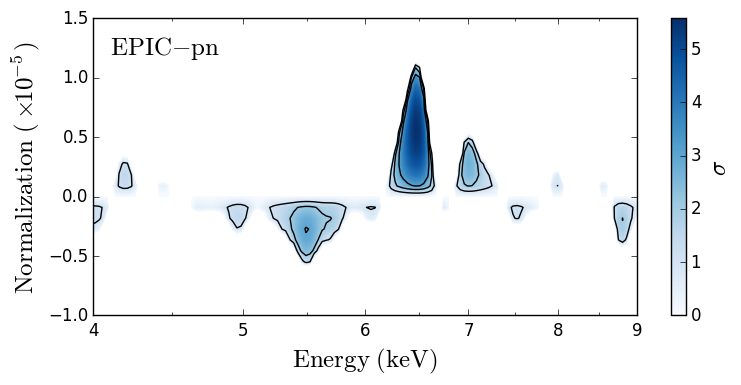}
\includegraphics[width=\linewidth]{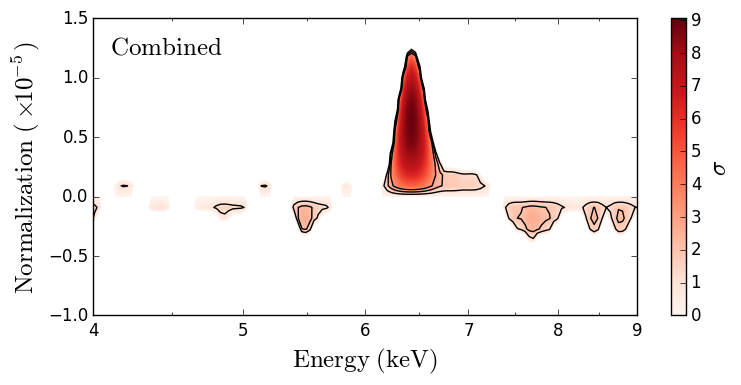}
\caption{Results of the search for emission and absorption lines, over the 4--9~keV band. The first plot shows the results for the EPIC-pn individually, and the combined plot shows the results from fitting all five instruments (pn, MOS1/2 and FPMA/B) simultaneously. Contour lines show the 1, 2 and 3$\sigma$ confidence levels. Normalizations are in units of photons~cm$^{-2}$~s$^{-1}$ and energy values are in the rest frame. The 7~keV line that appears in the EPIC-pn spectrum is not significant when the other data is taken into account.}
\label{fig_linesearch}
\end{figure}

From the EPIC-pn spectrum alone, it appears that there may be some additional complexity around the 6.4~keV iron line, most noticeably an apparent emission line at $\sim7$~keV. This could potentially be due to the iron K$\beta$ line, or could be due to an absorption feature superimposed on a broader emission line. To probe the iron line region in more detail, we use a simple line search over the 4--9~keV band, following the method outlined in \citet{Tombesi10} and assuming a power law continuum. The results are shown in Fig.~\ref{fig_linesearch}. We perform the search separately for each set of detectors, then in a combined fit, so that detections can be compared between instruments. In all cases, we find the 6.4~keV line to be highly significant, with the only other feature significant at the 3$\sigma$ level being a spurious feature at $\sim 5.5$~keV in the EPIC-pn, which does not correspond to any features in the other instruments. There is a general trend towards an excess at the high energy side of the iron line, which may correspond to presence of an additional line or lines, a broad component, or be due to the presence of a 7~keV absorption edge. There are various features significant at the 2$\sigma$ level above 7~keV in the combined analysis (including \nustar , and MOS), however they are not consistent in size or energy and most likely correspond to the continuum over-predicting the flux in this band \citep[we note that this method is extremely sensitive to the choice of baseline model;][]{Zoghbi15}. The absence of any apparent ionized absorption (either warm absorber or ultra-fast outflow) in this source is interesting, and may be relevant to the change in Seyfert classification. In particular, it implies that either there is a weak or absent disk wind or that we are observing the source at such an angle that the wind does not intercept the line of sight.

We fit the combined X-ray spectrum with a variety of models, with the aim of measuring the level of the intrinsic absorption and for easy comparison with other sources. The fit results are shown in Table~\ref{table_xrayfit}. In all fits we allow for a normalization offset between the 5 detectors. The following models are used:
\begin{itemize}
\item Model 1: Fitting \xmm\ data from 0.3--2~keV with an absorbed (Galactic and intrinsic absorption) power law. This model is for comparison with the \rosat\ and \xmm\ slew surveys, which fit models over similar bands.
\item Model 2: As model 1, but fit up to 10~keV, again only using \xmm\ data and including a narrow emission line at 6.4~keV. This model allows for more general comparison with \xmm\ (and other) observed AGN.
\item Model 3: As model 2, but now including a soft excess, modelled with a black body (\emph{bbody} in \textsc{xspec}). We test adding both neutral (modelled using \emph{zpcfabs}) and ionized \citep[modelled using \emph{zxipcf},][]{Reeves08} partially-covering absorbers, and find no improvement to the fit from adding either.
\item Model 4: As model 3, but now including the \nustar\ data from 3--79~keV, and replacing the power law with a cut-off power law (\emph{cutoffpl}). Again, we try including additional partial covering absorption and find no improvement to the fit. This is our best-fit phenomenological model, which includes spectral components to describe all major features of the spectrum.
\item Model 5: A more physical model, where we replace the narrow Gaussian line with a distant reflection component modelled with \emph{xillver} \citep{Garcia10}, and use a combination of Comptonization \citep[\emph{comptt};][]{Titarchuk94} and relativistic reflection \citep[\emph{relxill};][]{Garcia14} to fit the soft excess and broad-band spectral curvature. We use a broken power law emissivity profile, fixing the outer index at 3, and the break radius at 6~$R_\textrm{G}$. Removing either of the reflection and Comptonization components worsens the fit significantly ($\Delta\chi^2\sim50$), although it is not entirely clear what spectral features this is due to in the reflection case. It is most likely required to fit subtle spectral curvature in the high-quality broad-band spectrum. As with the other models, additional absorbers do not improve the fit significantly.
\end{itemize}

\begin{table}
\centering
\caption{X-ray spectral fit parameters for the different models. Models 1--4 are phenomenological, and model 5 is made up of physical components. Models 1--3 are fit to different subsets of the data.}
\label{table_xrayfit}
\begin{tabular}{l l l l}
\hline
\hline
Parameter & Value & Units & Description\\
\hline
\multicolumn{4}{l}{Model 1 - \emph{tbabs$\times$ztbabs$\times$(powerlaw)} - (0.3--2~keV)}\\
$n_\textrm{H}$ & $1.50\pm0.03$ & $10^{21}$ cm$^{-2}$ & Column density\\
$\Gamma$ & $2.00\pm0.02$ & & Power law index\\
$\chi^2/\textrm{d.o.f.}$ &282/150=1.88&\\
\hline

\multicolumn{4}{l}{Model 2 - \emph{tbabs$\times$ztbabs$\times$(powerlaw+zgauss)} - (0.3--10~keV)}\\
$n_\textrm{H}$ & $1.09\pm0.01$ & $10^{21}$ cm$^{-2}$ & Column density\\
$\Gamma$ & $1.764\pm0.004$ & & Power law index\\
$E_\textrm{Gauss}$ & $6.421\pm0.003$ & keV&Line energy\\
$\chi^2/\textrm{d.o.f.}$ &1468/504=2.91\\

\hline

\multicolumn{4}{l}{Model 3 - \emph{tbabs$\times$ztbabs$\times$(bbody+powerlaw+zgauss)}} \\ 
(0.3--10~keV)\\
$n_\textrm{H}$ & $0.95\pm0.03$ & $10^{21}$ cm$^{-2}$ & Column density\\
$\Gamma$ & $1.60\pm0.01$ & & Power law index\\
$E_\textrm{Gauss}$ & $6.44\pm0.01$ &keV& Line energy\\
$kT$ & $0.24\pm0.01$ & keV & Black body temp.\\
$\chi^2/\textrm{d.o.f.}$ &764/502=1.52&\\

\hline

\multicolumn{4}{l}{Model 4 - \emph{tbabs$\times$ztbabs$\times$(bbody+cutoffpl+zgauss)}} \\ 
(0.3--79~keV)\\
$n_\textrm{H}$ & $0.94\pm0.03$ & $10^{21}$ cm$^{-2}$ & Column density\\
$\Gamma$ & $1.57\pm0.02$ & & Power law index\\
$E_\textrm{cut}$ & $102_{-16}^{+23}$ & keV & Cut-off energy\\
$E_\textrm{Gauss}$ & $6.43\pm0.01$ &keV& Line energy\\
$kT$ & $0.24\pm0.01$ & keV & Black body temp.\\

$\chi^2/\textrm{d.o.f.}$ &1144/851=1.34&\\

\hline

\multicolumn{4}{l}{Model 5 - \emph{tbabs$\times$ztbabs$\times$(comptt+cutoffpl+xillver+relxill)}} \\ 
(0.3--79~keV)\\
$n_\textrm{H}$ & $1.55\pm0.01$ & $10^{21}$ cm$^{-2}$ & Column density\\
$\Gamma$ & $<1.43$ & & Power law index\\
$E_\textrm{cut}$ & $40_{-2}^{+6}$ & keV & Cut-off energy\\
$T_0$ & $0.93_{-0.03}^{+0.02}$ & keV & Seed temp.\\
$kT$ & $<1.9$ & keV & Plasma temp.\\
$\tau$ & $6.7_{-0.4}^{+0.2}$ & & Optical depth\\
$A_\textrm{Fe}$ & $>9$ & & Iron abundance\\
$q_\textrm{in}$ & $>8.5$ & & Emissivity index\\
$i$ & $78_{01}^{+2}$ & degrees & Inclination\\
$a$ & $>0.995$ & & Spin\\
$\log(\xi)$ & $2.4\pm0.1$ & erg cm s$^{-1}$& Ionization\\

$\chi^2/\textrm{d.o.f.}$ &1028/841=1.22&\\

\hline
\multicolumn{4}{l}{Cross-normalization constants (relative to EPIC-pn)}\\
$C_\textrm{MOS1}$ & $0.859\pm0.003$ \\
$C_\textrm{MOS2}$ & $1.036\pm0.004$ \\
$C_\textrm{FPMA}$ & $1.06\pm0.01$ \\
$C_\textrm{FPMB}$ & $1.09\pm0.01$ \\

\hline
\multicolumn{4}{l}{Luminosities (EPIC-pn)}\\
$L$(0.5--2) & $1.15\times10^{43}$ & erg s$^{-1}$ \\
$L$(2--10) & $1.69\times10^{43}$ & erg s$^{-1}$ \\

\hline
\hline

\end{tabular}
Luminosities and normalization constants are calculated from model 5, and luminosities are corrected for absorption (both Galactic and intrinsic).
\end{table}

\begin{figure}
\centering
\includegraphics[width=\linewidth]{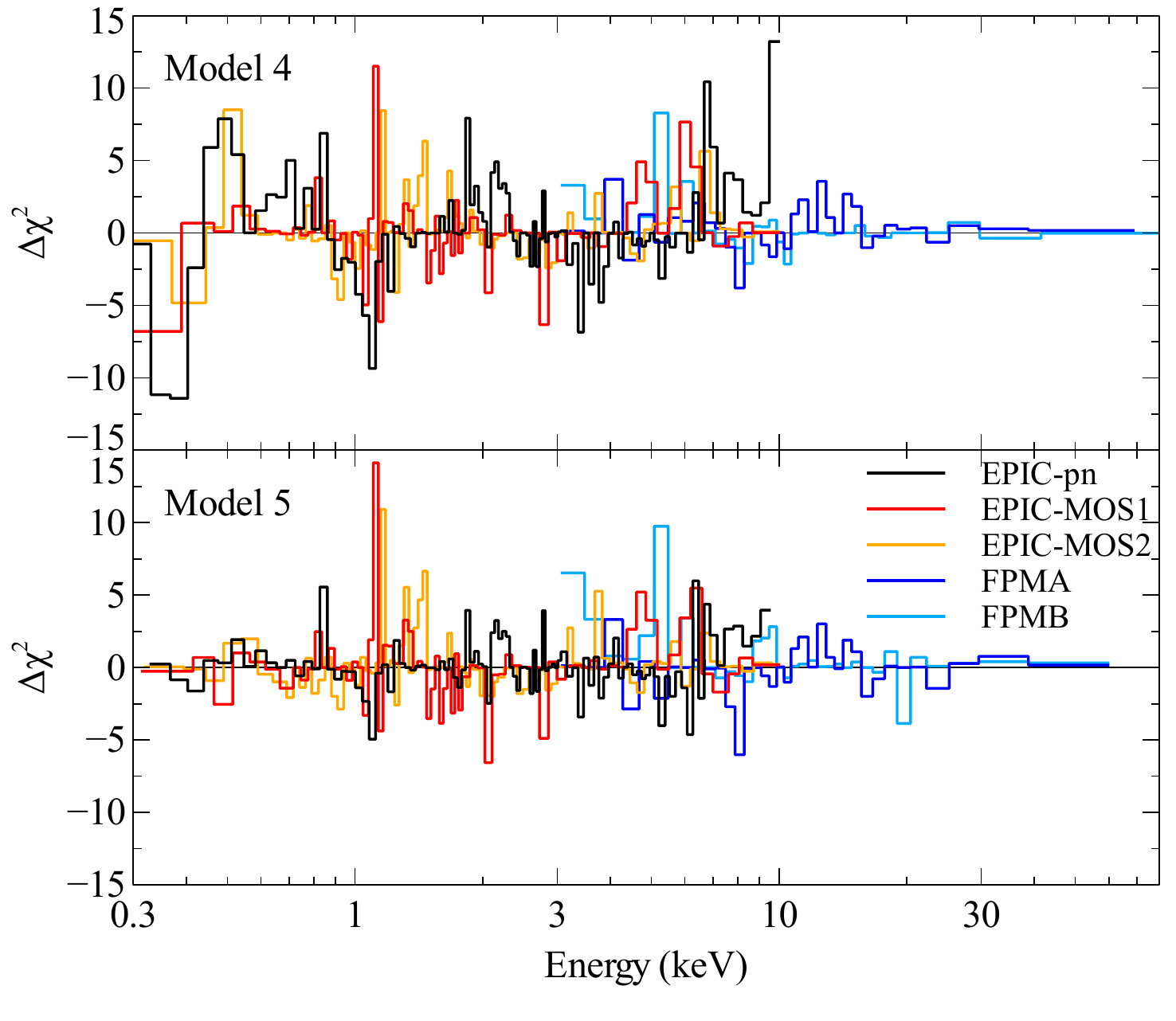}
\caption{$\chi^2$ residuals for the best-fit phenomenological and physical models (top and bottom, respectively). Different colours correspond to the five different instruments. A disagreement between the MOS and pn detectors is seen just above 1~keV, and a slightly different slope is found between the FPMs and the EPIC instruments between 3--10~keV. Aside from these instrumental effects, there are no significant residuals in the physical model. The phenomenological model shows strong residuals around the soft excess. The corresponding models are shown in Fig.~\ref{fig_model}}
\label{fig_broadband_resids}
\end{figure}

\begin{figure*}
\centering
\includegraphics[width=0.8\linewidth]{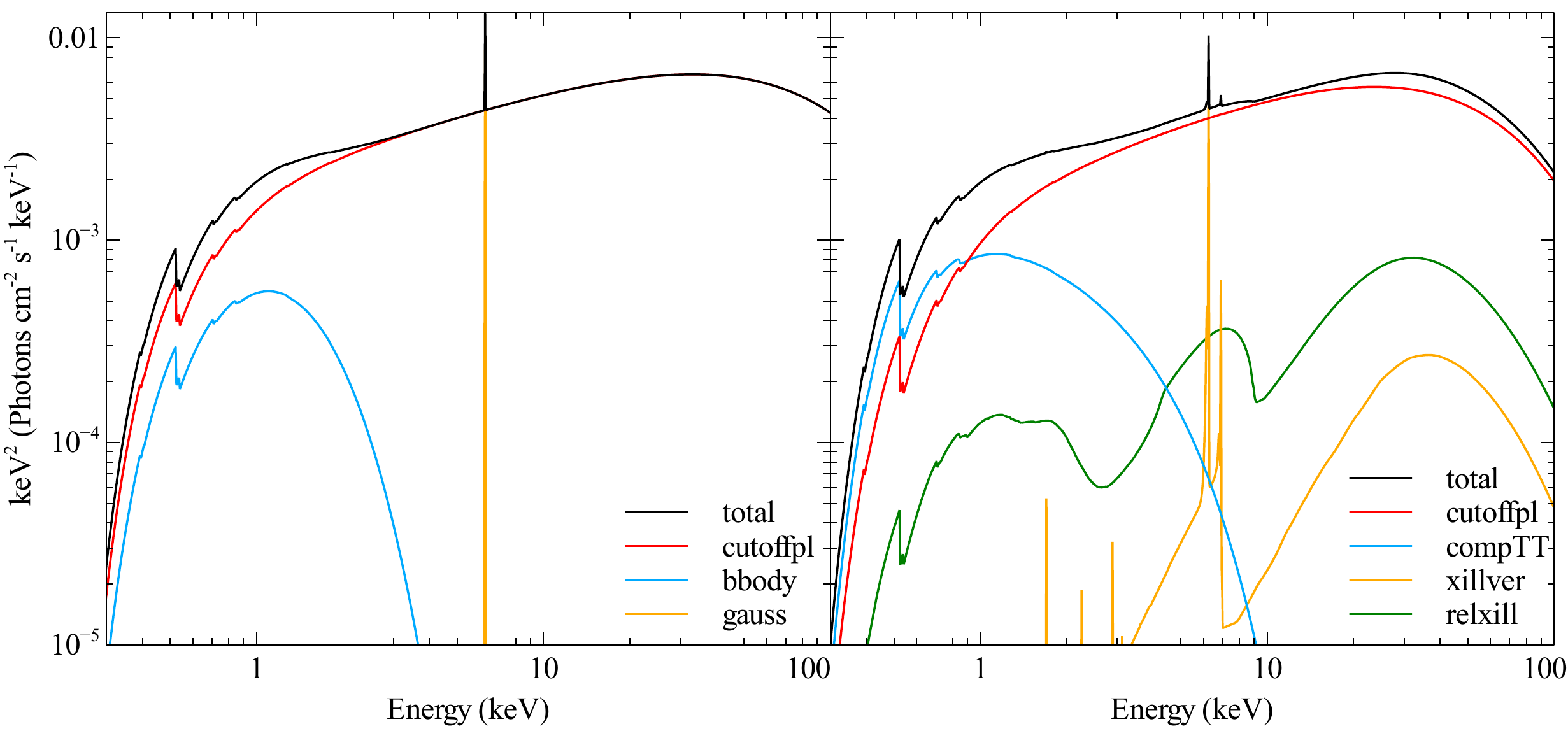}
\caption{Best-fit broad-band spectral models, showing the different spectral components used. Left and right plots correspond to the phenomenological and physical models, respectively. In both cases, the spectrum is dominated by the continuum power-law, with a strong narrow iron line.}
\label{fig_model}
\end{figure*}

Models 1--4 are phenomenological, and are intended for comparison purposes. We show the residuals to the best-fit models 4 and 5 in Fig.~\ref{fig_broadband_resids}, and the models themselves in Fig.~\ref{fig_model}. For model 4, the main residual features are at low energies, and presumably arise from our phenomenological model of the soft excess. There is a small disagreement in $\Gamma$ between the \xmm\ and \nustar\ instruments from 3--10~keV, which also contributes to the relatively high $\chi^2$ values we find. For model 5, the main residuals are around 1~keV, where there is a disagreement between the MOS and pn detectors, and from 3--10~keV, where the \nustar\ and \xmm\ spectra overlap.

\begin{figure}
\centering
\includegraphics[width=\linewidth]{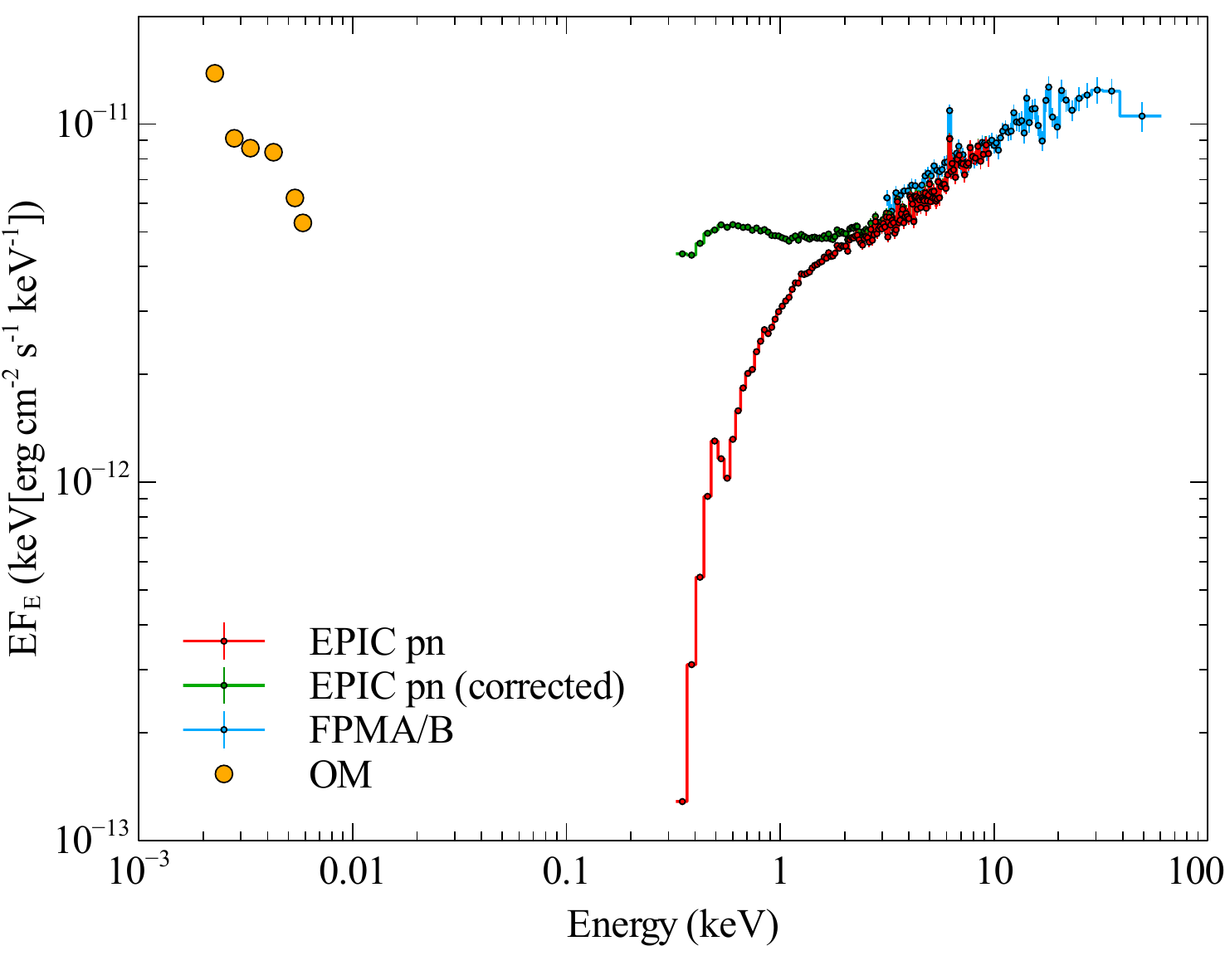}
\caption{\xmm\ and \nustar\ SED. Green points show the EPIC-pn spectrum after correcting for both Galactic and intrinsic absorption.}
\label{fig_sed}
\end{figure}

We can estimate from the best fit model (model 5) the column density that would be needed to reduce the X-ray flux to the level observed in 1990. By adjusting the column density of the neutral absorber at the source redshift (\emph{ztbabs}), we find that the column density needed is $\sim2.45\times10^{22}$~cm$^{-2}$, compared to the observed $1.5\times10^{21}$~cm$^{-2}$ in 2014. However, the column density measured in the 2014 observations is not sufficient to explain the rise in flux to the highest flux \swift\ XRT observation. The XRT maximum 0.2--2~keV flux is more than twice that found with \xmm , and removing the absorber completely only produces a ~50\% increase in flux. This suggests that the majority of the short term X-ray variability is intrinsic to the source.

In Fig.~\ref{fig_sed} we show the SED of HE\,1136-2304 from 2014-07-02 with the \xmm\ EPIC-pn and optical monitor (OM), and \nustar. We also show the effect of the absorption, by correcting the EPIC-pn spectrum for both intrinsic and Galactic absorption. It is clear from this that a large fraction of the flux is being emitted in the X-rays.

\subsection{Optical spectrum}

\begin{figure*}
  \centering
    \includegraphics[width=12cm,angle=270]{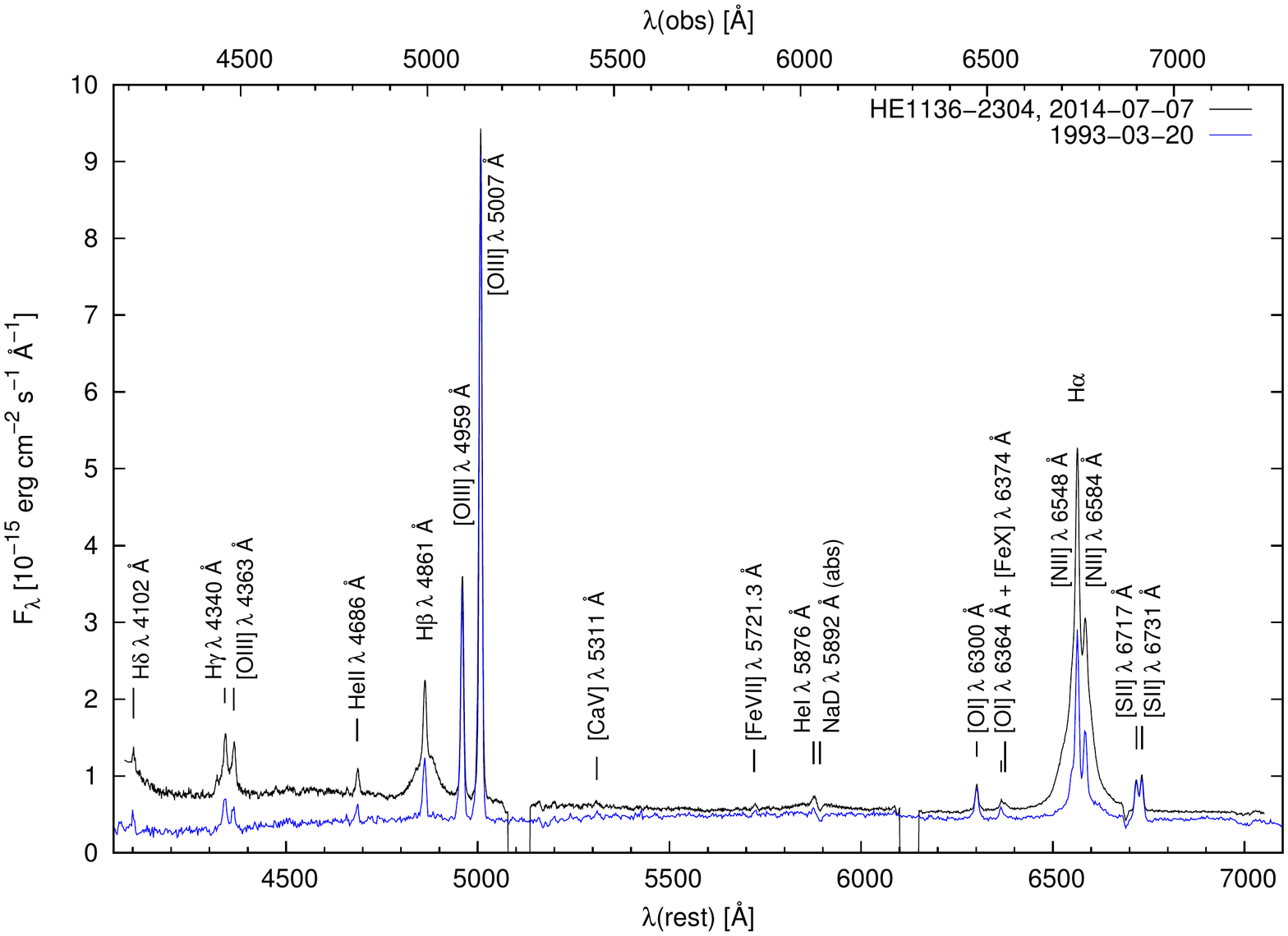}
      \caption{Optical spectrum of HE1136-2304 taken with the SALT telescope
on July 7th, 2014 (black line).
A further spectrum of this AGN taken in 1993 (blue line) is overlayed
for comparison. The broad H$\alpha$ and H$\beta$ lines are clearly much stronger in the 2014 spectrum than in the 1993 spectrum.}
       \vspace*{-4mm} 
         \label{HE1136-2304_SALT_2014-07-07.ps}
   \end{figure*}

The optical spectrum of HE1136-2304 taken with the SALT telescope in 2014 is presented
in Fig.~\ref{HE1136-2304_SALT_2014-07-07.ps}. This spectrum is shown in the rest frame of the AGN. Based on this observed spectrum we calculate a blue magnitude m$_{B}$ of 17.3$\pm{}$0.1 for HE\,1136-2304
%

We derived the flux of the emission line intensities by integrating the flux above a linearly interpolated continuum, locally defined by regions just blueward and redward of the individual emission lines. The fluxes were integrated for the wavelength ranges given in column 3. We present in Table~\ref{table_opticallines} the observed emission line intensities as well as those corrected for Galactic extinction\footnote{Note that we have neglected a possible weak \ion{Fe}{ii} component in our optical spectra which will overlap with the [\ion{O}{iii}] and H$\beta$ lines. Given how bright these lines are relative to the \ion{Fe}{ii} component, this contamination is negligible.
Additionally, the relative flux contribution of the [\ion{N}{ii}] lines with respect to the H$\alpha$-[\ion{N}{ii}] line complex is far stronger in the faint state, introducing additional uncertainty into the measurements of the H$\alpha$ fluxes.}.
With respect to the relative intensities of the broad and narrow Balmer line components, for example H$\beta$,  HE\,1136-2304 has to be classified as Seyfert 1.5 type based on its spectrum taken in July 2014.  This is based on the definition of \citet{osterbrock06} that Seyfert galaxies 
with intermediate-type HI profiles in which both components can easily be recognized
are of Seyfert type 1.5. 
\begin{table}
\centering
\tabcolsep+1mm
\caption{Rest-frame optical emission line intensities of HE1136-2304 from 2014:
  observed values (2) and corrected for Galactic extinction (3).}
\label{table_opticallines}
\begin{tabular}{lccc}
\hline
\noalign{\smallskip}
Emission line                 & \multicolumn{2}{c}{Flux} & Wavelength range \\
                              & obs.  &  corr.        &   \\
\noalign{\smallskip}
(1)                           & (2)   & (3)           &    (4) \\
\noalign{\smallskip}
\hline
\noalign{\smallskip}
H$\gamma$                                      &   20$\pm{}$2     & 24.5$\pm{}$2    & 4311 -- 4386 \\
$[$\ion{O}{iii}$]\,\lambda 4363$               &  5.5$\pm{}$1     & 6.2$\pm{}$1    & 4353 -- 4374 \\
\ion{He}{ii}\,$\lambda 4686$                   & 3.3$\pm{}$0.1    & 4.3$\pm{}$0.1   & 4674 -- 4698 \\
H$\beta$  (broad)                              &   40$\pm{}$5     & 43$\pm{}$5    & 4782 -- 4936 \\
H$\beta$  (narrow)                             &    9$\pm{}$2     & 10$\pm{}$2    & 4850 -- 4871 \\
$[$\ion{O}{iii}$]\lambda 4959$                 &   27$\pm{}$2     & 31$\pm{}$2    & 4940 -- 4979 \\
$[$\ion{O}{iii}$]$\,$\lambda 5007$             &   83$\pm{}$3     & 91$\pm{}$3    & 4984 -- 5027 \\
$[$\ion{Ca}{v}$]\lambda 5311$                  &  0.7$\pm{}$0.2   & 0.8$\pm{}$0.2  & 5300 -- 5321 \\
$[$\ion{Fe}{vii}$]\lambda 5721$                &  0.8$\pm{}$0.1   & 0.9$\pm{}$0.1  & 5712 -- 5733 \\
\ion{He}{i}$\lambda 5876$                      &  1.4$\pm{}$0.1   & 1.6$\pm{}$0.1  & 5864 -- 5887 \\
$[$\ion{O}{i}$]\lambda 6300$                   &  3.9$\pm{}$0.5   & 4.2$\pm{}$0.5  & 6279 -- 6320 \\
$[$\ion{O}{i}$]\lambda 6364$                   &  1.4$\pm{}$0.3   & 1.5$\pm{}$0.3  & 6348 -- 6373 \\
$[$\ion{Fe}{X}$]\lambda 6374$                  & 0.93$\pm{}$0.2   & 0.95$\pm{}$0.3  & 6373 -- 6396 \\
H$\alpha$ (broad, incl.[$\ion{N}{ii}$])        &  191$\pm{}$25    & 211$\pm{}$25   & 6400 -- 6670 \\
H$\alpha$ (narrow)                             &   20$\pm{}$2     & 22$\pm{}$2    & 6556 -- 6572 \\
$[$\ion{N}{ii}$]\lambda 6584$                  &  7.6$\pm{}$1     & 7.8$\pm{}$1    & 6576 -- 6598 \\
$[$\ion{S}{ii}$]\lambda 6717$                  &   4.3$\pm{}$0.4  & 4.4$\pm{}$0.5  & 6706 -- 6725 \\
$[$\ion{S}{ii}$]\lambda 6731$                  &   4.5$\pm{}$0.4  & 5.0$\pm{}$0.5  & 6725 -- 6745 \\
\noalign{\smallskip}
\hline
\end{tabular}\\
Line fluxes (2) in units of 10$^{-15}$\,erg\,s$^{-1}$\,cm$^{-2}$.
\end{table}

The relative intensities of the highly ionized lines are quite strong. The [\ion{Fe}{x}]$\lambda$6374 line intensity, for example, is stronger in HE\,1136-2304 than in the prototype Seyfert 1.5 galaxy NGC~5548 \citep{peterson94}.
The narrow lines, for example [\ion{O}{iii}]$\lambda$5007, hold line widths  (FWHM) of 510 $\pm{}$ 10 \kms. The broad H$\beta$ component shows a width of 4200 $\pm{}$ 200 \kms.  These values are typical
for Seyfert galaxies. We determined the line widths in units of \kms\  by converting our spectrum 
into velocity space with the IRAF task `disptrans'.
This broad component of H$\beta$ is redshifted with respect to the narrow component by 250$\pm{}$50 \kms.

\begin{figure}
\centering
\includegraphics[width=\linewidth]{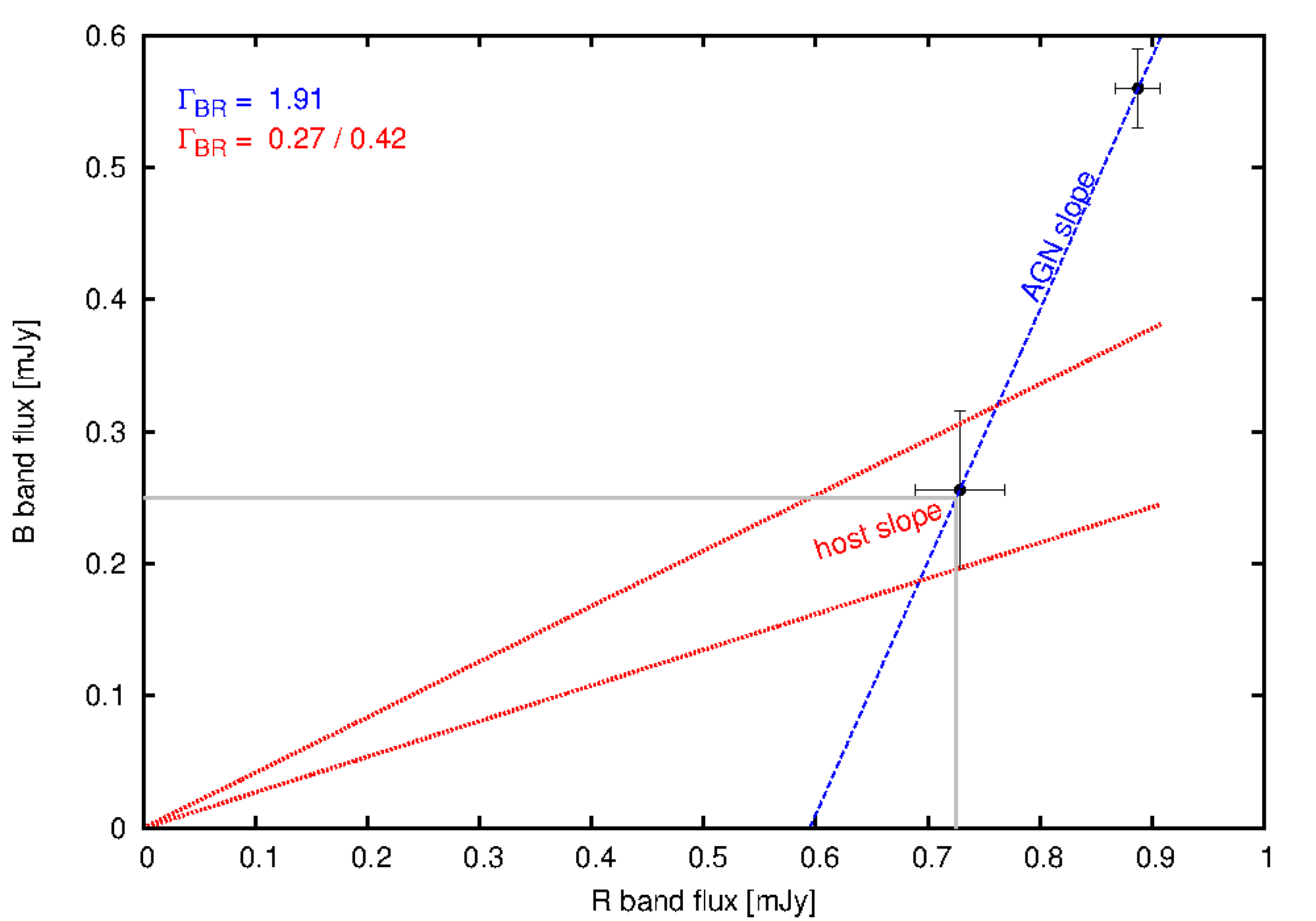}
\caption{B versus R flux variations of He1136-2304. The two measurements of He1136-2304
in the bright state (2014) and in the low state (1993) are used to determine
the AGN slope (blue). The red dashed lines give the range of host slopes for
nearby AGN as determined by \citet{Sakata10}. The intersections between the 
AGN and host galaxy slopes gives the possible range of host galaxy fluxes in the B and R bands (grey lines indicate the central value of R$=0.73$~mJy, B$=0.25$~mJy).}
\label{fig_fvgdiagram}
\end{figure}

We use the flux variation gradient (FVG) method \citep{Choloniewski81,Winkler92,Nunoz15} to estimate the relative contributions of the host and AGN to the continuum flux (shown in Fig.~\ref{fig_fvgdiagram}). We calculate B and R-band fluxes for each spectrum by convolving them with B and R filter curves. From 1993 to 2014 the R-band flux increases from $0.73\pm0.04$ to $0.89\pm0.02$ and the B-band from $0.26\pm0.06$ to $0.56\pm0.03$. The AGN and host flux lines intercept at R and B band fluxes of 0.73~mJy and 0.25~mJy, respectively. From this, we can infer that the maximum contribution of the AGN to the blue continuum in 1993 was $\sim$20\%, and that it must have increased by a factor of at least 4 in 2014. Similarly, the 1993 broad H$\alpha$ and H$\beta$ line fluxes were $(55\pm10)\times10^{-15}$\,erg\,s$^{-1}$\,cm$^{-2}$ and $(15\pm5) \times10^{-15}$\,erg\,s$^{-1}$\,cm$^{-2}$, a factor of 3--4 below those measured in 2014. A caveat here is that the two spectra were taken with different aperture sizes, so may contain slightly different host galaxy contributions.

Based on the column density needed to lower the X-ray flux to the level of 1990 ($\sim2.45\times10^{22}$~cm$^{-2}$, see \S~\ref{section_xray_results}), we can estimate the corresponding effect that the associated reddening should have had on the blue continuum flux, in the case that a drop in the absorbing column is the cause of the X-ray flare. Converting the column density to extinction using the relation from \citet{Predehl95} and assuming $A_V=3.1E_{B-V}$, we find that the flux at 5500~\AA\ should be lower by a factor of $\sim10^5$ with respect to the 2014 flux. We note that in AGN the ratio of $A_V$ to $N_\textrm{H}$ is generally significantly lower than that from Galactic absorption: \citet{Maiolino01} find that most AGN have $A_V/N_\textrm{H}$ a factor of 10 lower than Galactic, with some up to 100 times lower. Lowering the $A_V$ values by 10 and 100 corresponds to flux ratios ($F(2014)/F(1993)$) of 46 and 1.5, respectively.

The high-amplitude variability of HE\,1136-2304 makes BH mass estimates more uncertain. For a first order of magnitude estimate, we use the optical continuum luminosity, which varied very little. Using the width of the broad H$\beta$ line, continuum luminosity ($\log[\lambda L_\lambda (5100\textrm{\AA})]=42.72\pm0.05$) and the mass scaling relation of \citet[][equation 5]{Vestergaard06} we find a black hole mass of $\log(M_\textrm{BH}/M_\odot)=7.3\pm0.4$ in 2014 and $6.9\pm0.5$ in 1993, assuming 50\% and 10\% of the luminosity at 5100{\AA} is due to the AGN, respectively, based on the FVG results. 
Assuming a bolometric luminosity of $L_\textrm{bol}\approx9\lambda L_\lambda (5100\textrm{\AA})=2.3\times10^{43}$~erg s$^{-1}$ in 2014 \citep{Kaspi00} we find an Eddington fraction of $\sim0.01$. This may be an underestimate, however, as the 0.5--10~keV X-ray luminosity is $\sim2.8\times10^{43}$~erg s$^{-1}$, which would suggest the bolometric luminosity should be significantly larger.
From the 2--10~keV luminosity (Table~\ref{table_xrayfit}) we can calculate an alternative estimate of the bolometric luminosity of $3.4\times10^{44}$~erg s$^{-1}$, assuming a bolometric correction factor of $\kappa_\textrm{2--10~keV}=20$ \citep[see ][]{Vasudevan09}. This corresponds to an Eddington fraction of $\sim0.09$, which is significantly higher than that found from the continuum luminosity. 
Note that this range of inferred Eddington fractions is on the
same order as the critical accretion rate ($\dot{m}/\dot{m}_\textrm{Edd}=0.01$)
required in the model of \citet{Nicastro00} for the onset of a significant BLR. 
As seen in the SED (Fig.~\ref{fig_sed}) the X-ray flux is relatively high, which explains the difference in results found here.

For an independent estimate of the historical X-ray emission, we have used the [\ion{O}{iii}] emission from the narrow-line region Table~\ref{table_opticallines}. Based on the correlation between [\ion{O}{iii}] luminosity and (2-10) keV X-ray luminosity of \citet{Lamastra09}, and the observed [\ion{O}{iii}] luminosity of HE\,1136-2304, we predict an X-ray luminosity of 1.2 $\times 10^{42}$ erg~s$^{-1}$, which is a factor of 14 lower than the one measured during the deep XMM-Newton observation (Table~\ref{table_xrayfit})\footnote{Note that we did not apply any extinction correction when estimating $L_\textrm{[\ion{O}{iii}]}$. While we cannot estimate the BLR reddening from the optical spectrum because broad H$\alpha$ is blended with [NII], the narrow Balmer-line ratio does not indicate any NLR reddening.}. This provides independent evidence that HE\,1136-2304 is currently undergoing an epoch of enhanced X-ray emission.

\section{Discussion}
\label{section_discussion}

While the 2014 optical and X-ray data are simultaneous within a few days, an important caveat for the following discussion is that the earlier archival data are not simultaneous. There is a three year gap between the two, which is long enough for dramatic changes to occur in both the X-ray and optical bands, meaning that the observed fluxes and spectra are not necessarily identical to those in the corresponding band at the time of the other observation.

We fit the X-ray spectra with a variety of phenomenological models, all of which agree that 
a significant absorbing column (around $10^{21}$~cm$^{-2}$) is required in addition to 
Galactic absorption, although in no case does it greatly exceed this value. The spectrum is fairly conventional for a Seyfert 1.5. It shows a moderate soft excess, a narrow iron line at 6.4~keV and a high energy cut-off at $\sim100$~keV \citep[\nustar\ has been shown to be capable of accurately constraining cut-off energies outside the energy band; e.g.][]{Matt15, Garcia15}. This cut-off energy is fairly low, but not exceptional \citep{Fabian15}.
Based on the two EPIC-pn spectra shown in the right panel of Fig.~\ref{fig_lightcurves_comparison}, 
it is unlikely that absorption variability can be responsible for all of the variations in flux 
seen, as the $\sim30$~\% drop in flux seen by \xmm\ appears to be uniform across the bandpass, 
inconsistent with a change in the absorption column. Notably, there is a lack of ionized absorption in the form of either a warm absorber or an outflow, which may indicate a viewing angle that prevents the line of sight being intercepted by this material.

While the physical model gives the best overall fit to the X-ray spectrum, we caution that there are problems with this model. Several of the parameters are hitting their limits, particularly in the reflection model. Since no broad iron line is apparent in the spectrum it is difficult to be certain about the measured values. While the reflection component greatly improves the fit quality this may be for more complex reasons, such as a more complex continuum model being needed to properly reproduce the data.

Using SALT spectroscopy we have determined that HE\,1136-2304 changed its spectral 
type from nearly Seyfert 2 type (i.e. Seyfert 1.95) in March 1993 to a Seyfert 1.5 type in July 2014, 
coinciding with a huge increase in X-ray flux measured with \xmm\ and \swift .  There was only a very weak 
broad H$\alpha$ component apparent in the optical spectrum 
taken in March 1993
\citep[see optical spectrum in][our Fig.~\ref{HE1136-2304_SALT_2014-07-07.ps}]{Reimers96} and
there was no indication for a broad H$\beta$ component in 1993.
The definition for a Seyfert 1.9 type is based on that of \citet{osterbrock06}, i.e. that no broad H$\beta$ emission can be seen in the spectrum. We scaled the spectrum taken in the year 1993 with respect to the intensity of the [\ion{O}{iii}]\,$\lambda$5007 line taken in 2014. 

In contrast to the Balmer line and X-ray flux we did not see much change in the optical continuum flux when comparing the spectra taken in 1993 and 2014. 
While the general optical continuum flux strength is similar at both epochs there is a difference in the gradient, which is bluer in 2014 relative to 1993. We give in Table~5 the derived continuum intensities at three optical continuum wavelengths for the years 1993 and 2014. 
%

\begin{table}
\centering
\caption{ Optical continuum fluxes for the years 1993 and 2014.}
\begin{tabular}{ccc}
\noalign{\smallskip}
\hline
\noalign{\smallskip}
Wavelength [\AA{}]    & Flux (1993) & Flux (2014)\\
                     & \multicolumn{2}{c}{[10$^{-15}$\,erg\,s$^{-1}$\,cm$^{-2}$\,\AA$^{-1}$]}\\
\noalign{\smallskip}
\hline
\noalign{\smallskip}
4400 &  0.32$\pm{}$0.03&  0.77$\pm{}$ 0.03  \\
5500 &    0.48$\pm{}$0.02&  0.57$\pm{}$ 0.01  \\
6950 &    0.44$\pm{}$0.01&  0.53$\pm{}$ 0.01 \\
\noalign{\smallskip}
\hline
\noalign{\smallskip}
\end{tabular}
\end{table}

There are several potential explanations for the observed increase in X-ray flux and appearance of broad lines, 
which we will now consider:
\begin{itemize}
\item Some \emph{inactive} galaxies have been seen to flare due to stellar tidal disruption 
events \citep[TDEs; e.g.,][]{Komossa99}. HE\,1136-2304 itself is a classical AGN, judged from
its optical emission line ratios ($\log$ ([\ion{O}{iii}]5007/H$\beta$) = 0.96,  
$\log$ ([\ion{S}{ii}]6724/H$\beta$) = --0.35) which place it well within the AGN regime in diagnostic diagrams.
While TDEs can also occur in AGN, it is more difficult to make a positive case for them in systems
which permanently harbour an accretion disk, as the increase in flux is less dramatic relative to the non-flaring flux in an AGN. Further, we note that the high-state X-ray spectrum below
2 keV (Tab. 2), is rather flat, unlike the majority of soft X-ray TDEs which showed 
ultra-soft X-ray spectra \citep[e.g.][]{Bade96}. In addition, the detection of the source in \xmm\ slews from 2010 and 2011 is inconsitent with a single TDE, which would follow a well established decay profile on shorter timescales \citep{Rees88,Rees90}.

\item Another idea is that there is in fact no change in the intrinsic brightness -- instead, 
the broad lines and X-ray emission were obscured by dusty absorbing material, which left the line 
of sight, causing the apparent X-ray outburst and revealing the broad line region. The presence of mild 
absorption in the X-ray spectrum may support this, and the high inclination suggested by the reflection model is also consistent with this picture. 
\item \citet{Nicastro00} suggest a model where the width of the broad lines is controlled 
by the accretion rate \citep[see also][who showed that the sequence of AGN from type 1 to type 2 is controlled by the bolometric luminosity]{Elitzur14}. In this model the BLR is associated with a disc wind, the radius of which increases with accretion rate. Thus, as the accretion rate increases the breadth of the lines decreases, as it is determined by the Keplerian velocity at the wind radius. 
Below a limiting accretion rate, however, no such wind is produced, as evaporation inhibits its formation. This model could therefore explain our observations if, as seems plausible, the accretion 
rate of HE\,1136-2304 has crossed this threshold value, causing the X-ray outburst and BLR 
simultaneously. 
\item Finally, arguably the simplest explanation is a flare in the emission from the EUV 
and X-rays due to an increase in the accretion rate, which then excited a larger amount of broad line emission. This can potentially be triggered by disk instabilities, without the need for any external cause \citep[e.g.][]{Honma91,Komossa99}.
In practice this model is difficult to distinguish from the previous scenario, but makes no 
assumptions on the nature of the BLR.
\end{itemize}


\section{Conclusions}
\label{section_conclusions}
We have presented X-ray and optical observations of the AGN HE\,1136-2304 with \xmm , \nustar , \swift , and SALT. The AGN was found to have increased in X-ray flux by a factor of $\sim30$, coinciding with the appearance of broad lines in the optical spectrum (and hence a change in classification from Seyfert 1.95 to Seyfert 1.5).

We find that the X-ray spectrum requires significant absorption in excess of the Galactic column, with a column density of $\sim10^{21}$~cm$^{-2}$. An increase in column density of approximately 1 order of magnitude (to 2--3$\times10^{22}$~cm$^{-2}$) would be enough to explain the lower flux observed in 1993. However, changes in the absorption cannot explain the increased flux seen by \swift , meaning the majority of the short-term variability must be intrinsic to the source. While this is not conclusive, as the long-term variability could be driven by a different mechanism, it favours the interpretation of the long-term change in flux as also intrinsic to the source, caused by an increase in the accretion rate.

Sources like HE\,1136-2304, which show simultaneous high-amplitude variability in X-rays and optical emission lines, provide us with tight constraints on the physics behind AGN classifications and accretion variability. If other sources where the optical continuum does not respond to the X-ray changes can be identified this may help us understand the nature of disk instabilities, particularly if the relevant timescales can be constrained. In general, radio-quiet AGN flaring in X-rays remain a relatively unexplored area, with great potential for shedding light on accretion physics.

\section*{Acknowledgements}
We thank the anonymous referee for their detailed and constructive feedback. We thank Lutz Wisotzki for making available the optical spectrum of HE1136-2304 taken in 1993, and we thank Vassilis Karamanavis for a careful reading of the manuscript. MLP acknowledges financial support from the Science and Technology Facilities Council (STFC). This paper is based on observations taken with the SALT telescope. This work has been supported by DFG grant Ko 857/32-2. Based on observations with \xmm , an ESA science mission with instruments and contributions directly funded by ESA Member States and NASA. This work made use of data from the \nustar\ mission, a project led by the California Institute of Technology, managed by the Jet Propulsion Laboratory, and funded by the National Aeronautics and Space Administration. This research has made use of the \nustar\ Data Analysis Software (NuSTARDAS) jointly developed by the ASI Science Data Center (ASDC, Italy) and the California Institute of Technology (USA). We would also like to thank Neil Gehrels for approving the \swift\ ToO requests, and the \swift\ science operation team for performing the observations.
\bibliographystyle{mnras}
\bibliography{bibliography_he1136}

\begin{thebibliography}{}
\makeatletter
\relax
\def\mn@urlcharsother{\let\do\@makeother \do\$\do\&\do\#\do\^\do\_\do\%\do\~}
\def\mn@doi{\begingroup\mn@urlcharsother \@ifnextchar [ {\mn@doi@}
  {\mn@doi@[]}}
\def\mn@doi@[#1]#2{\def\@tempa{#1}\ifx\@tempa\@empty \href
  {http://dx.doi.org/#2} {doi:#2}\else \href {http://dx.doi.org/#2} {#1}\fi
  \endgroup}
\def\mn@eprint#1#2{\mn@eprint@#1:#2::\@nil}
\def\mn@eprint@arXiv#1{\href {http://arxiv.org/abs/#1} {{\tt arXiv:#1}}}
\def\mn@eprint@dblp#1{\href {http://dblp.uni-trier.de/rec/bibtex/#1.xml}
  {dblp:#1}}
\def\mn@eprint@#1:#2:#3:#4\@nil{\def\@tempa {#1}\def\@tempb {#2}\def\@tempc
  {#3}\ifx \@tempc \@empty \let \@tempc \@tempb \let \@tempb \@tempa \fi \ifx
  \@tempb \@empty \def\@tempb {arXiv}\fi \@ifundefined
  {mn@eprint@\@tempb}{\@tempb:\@tempc}{\expandafter \expandafter \csname
  mn@eprint@\@tempb\endcsname \expandafter{\@tempc}}}

\bibitem[\protect\citeauthoryear{{Antonucci}}{{Antonucci}}{1993}]{Antonucci93}
{Antonucci} R.,  1993, \mn@doi [\araa] {10.1146/annurev.aa.31.090193.002353},
  \href {http://adsabs.harvard.edu/abs/1993ARA%26A..31..473A} {31, 473}

\bibitem[\protect\citeauthoryear{{Aretxaga}, {Joguet}, {Kunth}, {Melnick}  \&
  {Terlevich}}{{Aretxaga} et~al.}{1999}]{Aretxaga99}
{Aretxaga} I.,  {Joguet} B.,  {Kunth} D.,  {Melnick} J.,   {Terlevich} R.~J.,
  1999, \mn@doi [\apjl] {10.1086/312114}, \href
  {http://adsabs.harvard.edu/abs/1999ApJ...519L.123A} {519, L123}

\bibitem[\protect\citeauthoryear{{Bade}, {Komossa}  \& {Dahlem}}{{Bade}
  et~al.}{1996}]{Bade96}
{Bade} N.,  {Komossa} S.,   {Dahlem} M.,  1996, \aap, \href
  {http://adsabs.harvard.edu/abs/1996A%26A...309L..35B} {309, L35}

\bibitem[\protect\citeauthoryear{{Brandt}, {Pounds}  \& {Fink}}{{Brandt}
  et~al.}{1995}]{Brandt95}
{Brandt} W.~N.,  {Pounds} K.~A.,   {Fink} H.,  1995, \mnras, \href
  {http://adsabs.harvard.edu/abs/1995MNRAS.273L..47B} {273, L47}

\bibitem[\protect\citeauthoryear{{Choloniewski}}{{Choloniewski}}{1981}]{Choloniewski81}
{Choloniewski} J.,  1981, \actaa, \href
  {http://adsabs.harvard.edu/abs/1981AcA....31..293C} {31, 293}

\bibitem[\protect\citeauthoryear{{Cohen}, {Puetter}, {Rudy}, {Ake}  \&
  {Foltz}}{{Cohen} et~al.}{1986}]{Cohen86}
{Cohen} R.~D.,  {Puetter} R.~C.,  {Rudy} R.~J.,  {Ake} T.~B.,   {Foltz} C.~B.,
  1986, \mn@doi [\apj] {10.1086/164758}, \href
  {http://adsabs.harvard.edu/abs/1986ApJ...311..135C} {311, 135}

\bibitem[\protect\citeauthoryear{{Collin-Souffrin}, {Alloin}  \&
  {Andrillat}}{{Collin-Souffrin} et~al.}{1973}]{Collin-Souffrin73}
{Collin-Souffrin} S.,  {Alloin} D.,   {Andrillat} Y.,  1973, \aap, \href
  {http://adsabs.harvard.edu/abs/1973A%26A....22..343C} {22, 343}

\bibitem[\protect\citeauthoryear{{Condon}, {Cotton}, {Greisen}, {Yin},
  {Perley}, {Taylor}  \& {Broderick}}{{Condon} et~al.}{1998}]{Condon98}
{Condon} J.~J.,  {Cotton} W.~D.,  {Greisen} E.~W.,  {Yin} Q.~F.,  {Perley}
  R.~A.,  {Taylor} G.~B.,   {Broderick} J.~J.,  1998, \mn@doi [\aj]
  {10.1086/300337}, \href {http://adsabs.harvard.edu/abs/1998AJ....115.1693C}
  {115, 1693}

\bibitem[\protect\citeauthoryear{{Denney} et~al.,}{{Denney}
  et~al.}{2014}]{Denney14}
{Denney} K.~D.,  et~al., 2014, \mn@doi [\apj] {10.1088/0004-637X/796/2/134},
  \href {http://adsabs.harvard.edu/abs/2014ApJ...796..134D} {796, 134}

\bibitem[\protect\citeauthoryear{{Elitzur}, {Ho}  \& {Trump}}{{Elitzur}
  et~al.}{2014}]{Elitzur14}
{Elitzur} M.,  {Ho} L.~C.,   {Trump} J.~R.,  2014, \mn@doi [\mnras]
  {10.1093/mnras/stt2445}, \href
  {http://adsabs.harvard.edu/abs/2014MNRAS.438.3340E} {438, 3340}

\bibitem[\protect\citeauthoryear{{Fabian}, {Lohfink}, {Kara}, {Parker},
  {Vasudevan}  \& {Reynolds}}{{Fabian} et~al.}{2015}]{Fabian15}
{Fabian} A.~C.,  {Lohfink} A.,  {Kara} E.,  {Parker} M.~L.,  {Vasudevan} R.,
  {Reynolds} C.~S.,  2015, \mn@doi [\mnras] {10.1093/mnras/stv1218}, \href
  {http://adsabs.harvard.edu/abs/2015MNRAS.451.4375F} {451, 4375}

\bibitem[\protect\citeauthoryear{{Fitzpatrick}}{{Fitzpatrick}}{1999}]{fitzpatrick99}
{Fitzpatrick} E.~L.,  1999, \mn@doi [\pasp] {10.1086/316293}, \href
  {http://adsabs.harvard.edu/abs/1999PASP..111...63F} {111, 63}

\bibitem[\protect\citeauthoryear{{Gallo} et~al.,}{{Gallo}
  et~al.}{2014}]{Gallo14}
{Gallo} L.~C.,  et~al., 2014, preprint, \href
  {http://adsabs.harvard.edu/abs/2014arXiv1410.2330G} {} (\mn@eprint {arXiv}
  {1410.2330})

\bibitem[\protect\citeauthoryear{{Garc{\'{\i}}a} \& {Kallman}}{{Garc{\'{\i}}a}
  \& {Kallman}}{2010}]{Garcia10}
{Garc{\'{\i}}a} J.,  {Kallman} T.~R.,  2010, \mn@doi [\apj]
  {10.1088/0004-637X/718/2/695}, \href
  {http://adsabs.harvard.edu/abs/2010ApJ...718..695G} {718, 695}

\bibitem[\protect\citeauthoryear{{Garc{\'{\i}}a} et~al.,}{{Garc{\'{\i}}a}
  et~al.}{2014}]{Garcia14}
{Garc{\'{\i}}a} J.,  et~al., 2014, \mn@doi [\apj] {10.1088/0004-637X/782/2/76},
  \href {http://adsabs.harvard.edu/abs/2014ApJ...782...76G} {782, 76}

\bibitem[\protect\citeauthoryear{{Garcia}, {Dauser}, {Steiner}, {McClintock},
  {Keck}  \& {Wilms}}{{Garcia} et~al.}{2015}]{Garcia15}
{Garcia} J.~A.,  {Dauser} T.,  {Steiner} J.~F.,  {McClintock} J.~E.,  {Keck}
  M.~L.,   {Wilms} J.,  2015, preprint, \href
  {http://adsabs.harvard.edu/abs/2015arXiv150503616G} {} (\mn@eprint {arXiv}
  {1505.03616})

\bibitem[\protect\citeauthoryear{{Goodrich}}{{Goodrich}}{1989}]{Goodrich89}
{Goodrich} R.~W.,  1989, \mn@doi [\apj] {10.1086/167384}, \href
  {http://adsabs.harvard.edu/abs/1989ApJ...340..190G} {340, 190}

\bibitem[\protect\citeauthoryear{{Grupe}, {Beuermann}, {Mannheim}, {Bade},
  {Thomas}, {de Martino}  \& {Schwope}}{{Grupe} et~al.}{1995}]{Grupe95}
{Grupe} D.,  {Beuermann} K.,  {Mannheim} K.,  {Bade} N.,  {Thomas} H.-C.,  {de
  Martino} D.,   {Schwope} A.,  1995, \aap, \href
  {http://adsabs.harvard.edu/abs/1995A%26A...299L...5G} {299, L5}

\bibitem[\protect\citeauthoryear{{Grupe}, {Komossa}, {Gallo}, {Longinotti},
  {Fabian}, {Pradhan}, {Gruberbauer}  \& {Xu}}{{Grupe} et~al.}{2012}]{Grupe12}
{Grupe} D.,  {Komossa} S.,  {Gallo} L.~C.,  {Longinotti} A.~L.,  {Fabian}
  A.~C.,  {Pradhan} A.~K.,  {Gruberbauer} M.,   {Xu} D.,  2012, \mn@doi [\apjs]
  {10.1088/0067-0049/199/2/28}, \href
  {http://adsabs.harvard.edu/abs/2012ApJS..199...28G} {199, 28}

\bibitem[\protect\citeauthoryear{{Grupe}, {Komossa}, {Scharw{\"a}chter},
  {Dietrich}, {Leighly}, {Lucy}  \& {Barlow}}{{Grupe}
  et~al.}{2013}]{Grupe13_wpvs007}
{Grupe} D.,  {Komossa} S.,  {Scharw{\"a}chter} J.,  {Dietrich} M.,  {Leighly}
  K.~M.,  {Lucy} A.,   {Barlow} B.~N.,  2013, \mn@doi [\aj]
  {10.1088/0004-6256/146/4/78}, \href
  {http://adsabs.harvard.edu/abs/2013AJ....146...78G} {146, 78}

\bibitem[\protect\citeauthoryear{{Grupe}, {Komossa}  \& {Saxton}}{{Grupe}
  et~al.}{2015}]{Grupe15}
{Grupe} D.,  {Komossa} S.,   {Saxton} R.,  2015, \mn@doi [\apjl]
  {10.1088/2041-8205/803/2/L28}, \href
  {http://adsabs.harvard.edu/abs/2015ApJ...803L..28G} {803, L28}

\bibitem[\protect\citeauthoryear{{Guainazzi}}{{Guainazzi}}{2002}]{Guainazzi02_ngc6300}
{Guainazzi} M.,  2002, \mn@doi [\mnras] {10.1046/j.1365-8711.2002.05132.x},
  \href {http://adsabs.harvard.edu/abs/2002MNRAS.329L..13G} {329, L13}

\bibitem[\protect\citeauthoryear{{Guainazzi}, {Matt}, {Fiore}  \&
  {Perola}}{{Guainazzi} et~al.}{2002}]{Guainazzi02_ugc4203}
{Guainazzi} M.,  {Matt} G.,  {Fiore} F.,   {Perola} G.~C.,  2002, \mn@doi
  [\aap] {10.1051/0004-6361:20020471}, \href
  {http://adsabs.harvard.edu/abs/2002A%26A...388..787G} {388, 787}

\bibitem[\protect\citeauthoryear{{Harrison} et~al.,}{{Harrison}
  et~al.}{2013}]{Harrison13}
{Harrison} F.~A.,  et~al., 2013, \mn@doi [\apj] {10.1088/0004-637X/770/2/103},
  \href {http://adsabs.harvard.edu/abs/2013ApJ...770..103H} {770, 103}

\bibitem[\protect\citeauthoryear{{Honma}, {Matsumoto}  \& {Kato}}{{Honma}
  et~al.}{1991}]{Honma91}
{Honma} F.,  {Matsumoto} R.,   {Kato} S.,  1991, \pasj, \href
  {http://adsabs.harvard.edu/abs/1991PASJ...43..147H} {43, 147}

\bibitem[\protect\citeauthoryear{{Jansen} et~al.,}{{Jansen}
  et~al.}{2001}]{Jansen01}
{Jansen} F.,  et~al., 2001, \mn@doi [\aap] {10.1051/0004-6361:20000036}, \href
  {http://adsabs.harvard.edu/abs/2001A%26A...365L...1J} {365, L1}

\bibitem[\protect\citeauthoryear{{Kaspi}, {Smith}, {Netzer}, {Maoz}, {Jannuzi}
  \& {Giveon}}{{Kaspi} et~al.}{2000}]{Kaspi00}
{Kaspi} S.,  {Smith} P.~S.,  {Netzer} H.,  {Maoz} D.,  {Jannuzi} B.~T.,
  {Giveon} U.,  2000, \mn@doi [\apj] {10.1086/308704}, \href
  {http://adsabs.harvard.edu/abs/2000ApJ...533..631K} {533, 631}

\bibitem[\protect\citeauthoryear{{Kollatschny} \& {Fricke}}{{Kollatschny} \&
  {Fricke}}{1985}]{Kollatschny85}
{Kollatschny} W.,  {Fricke} K.~J.,  1985, \aap, \href
  {http://adsabs.harvard.edu/abs/1985A%26A...146L..11K} {146, L11}

\bibitem[\protect\citeauthoryear{{Kollatschny}, {Bischoff}, {Robinson}, {Welsh}
   \& {Hill}}{{Kollatschny} et~al.}{2001}]{kollatschny01}
{Kollatschny} W.,  {Bischoff} K.,  {Robinson} E.~L.,  {Welsh} W.~F.,   {Hill}
  G.~J.,  2001, \mn@doi [\aap] {10.1051/0004-6361:20011323}, \href
  {http://adsabs.harvard.edu/abs/2001A%26A...379..125K} {379, 125}

\bibitem[\protect\citeauthoryear{{Komossa}}{{Komossa}}{2015}]{Komossa15_TDEs}
{Komossa} S.,  2015, \mn@doi [Journal of High Energy Astrophysics]
  {10.1016/j.jheap.2015.04.006}, \href
  {http://esoads.eso.org/abs/2015JHEAp...7..148K} {7, 148}

\bibitem[\protect\citeauthoryear{{Komossa} \& {Bade}}{{Komossa} \&
  {Bade}}{1999}]{Komossa99}
{Komossa} S.,  {Bade} N.,  1999, \aap, \href
  {http://adsabs.harvard.edu/abs/1999A%26A...343..775K} {343, 775}

\bibitem[\protect\citeauthoryear{{Komossa} et~al.,}{{Komossa}
  et~al.}{2008}]{Komossa08}
{Komossa} S.,  et~al., 2008, \mn@doi [\apjl] {10.1086/588281}, \href
  {http://adsabs.harvard.edu/abs/2008ApJ...678L..13K} {678, L13}

\bibitem[\protect\citeauthoryear{{Komossa}, {Grupe}, {Saxton}  \&
  {Gallo}}{{Komossa} et~al.}{2014}]{Komossa15_seyferts}
{Komossa} S.,  {Grupe} D.,  {Saxton} R.,   {Gallo} L.,  2014, in Proceedings of
  Swift: 10 Years of Discovery (SWIFT 10), held 2-5 December 2014 at La
  Sapienza University, Rome, Italy. Online at <A
  href=''http://pos.sissa.it/cgi-bin/reader/conf.cgi?confid=233''>http://pos.sissa.it/cgi-bin/reader/conf.cgi?confid=233</A>,
  id.143. p.~143

\bibitem[\protect\citeauthoryear{{Korista} \& {Goad}}{{Korista} \&
  {Goad}}{2004}]{Korista04}
{Korista} K.~T.,  {Goad} M.~R.,  2004, \mn@doi [\apj] {10.1086/383193}, \href
  {http://adsabs.harvard.edu/abs/2004ApJ...606..749K} {606, 749}

\bibitem[\protect\citeauthoryear{{LaMassa} et~al.,}{{LaMassa}
  et~al.}{2015}]{LaMassa15}
{LaMassa} S.~M.,  et~al., 2015, \mn@doi [\apj] {10.1088/0004-637X/800/2/144},
  \href {http://adsabs.harvard.edu/abs/2015ApJ...800..144L} {800, 144}

\bibitem[\protect\citeauthoryear{{Lamastra}, {Bianchi}, {Matt}, {Perola},
  {Barcons}  \& {Carrera}}{{Lamastra} et~al.}{2009}]{Lamastra09}
{Lamastra} A.,  {Bianchi} S.,  {Matt} G.,  {Perola} G.~C.,  {Barcons} X.,
  {Carrera} F.~J.,  2009, \mn@doi [\aap] {10.1051/0004-6361/200912023}, \href
  {http://adsabs.harvard.edu/abs/2009A%26A...504...73L} {504, 73}

\bibitem[\protect\citeauthoryear{{Leighly}, {Cooper}, {Grupe}, {Terndrup}  \&
  {Komossa}}{{Leighly} et~al.}{2015}]{Leighly15}
{Leighly} K.~M.,  {Cooper} E.,  {Grupe} D.,  {Terndrup} D.~M.,   {Komossa} S.,
  2015, \mn@doi [\apjl] {10.1088/2041-8205/809/1/L13}, \href
  {http://adsabs.harvard.edu/abs/2015ApJ...809L..13L} {809, L13}

\bibitem[\protect\citeauthoryear{{MacLeod} et~al.,}{{MacLeod}
  et~al.}{2016}]{MacLeod16}
{MacLeod} C.~L.,  et~al., 2016, \mn@doi [\mnras] {10.1093/mnras/stv2997}, \href
  {http://adsabs.harvard.edu/abs/2016MNRAS.457..389M} {457, 389}

\bibitem[\protect\citeauthoryear{{Maiolino}, {Marconi}, {Salvati}, {Risaliti},
  {Severgnini}, {Oliva}, {La Franca}  \& {Vanzi}}{{Maiolino}
  et~al.}{2001}]{Maiolino01}
{Maiolino} R.,  {Marconi} A.,  {Salvati} M.,  {Risaliti} G.,  {Severgnini} P.,
  {Oliva} E.,  {La Franca} F.,   {Vanzi} L.,  2001, \mn@doi [\aap]
  {10.1051/0004-6361:20000177}, \href
  {http://adsabs.harvard.edu/abs/2001A%26A...365...28M} {365, 28}

\bibitem[\protect\citeauthoryear{{Marchese}, {Braito}, {Della Ceca},
  {Caccianiga}  \& {Severgnini}}{{Marchese} et~al.}{2012}]{Marchese12}
{Marchese} E.,  {Braito} V.,  {Della Ceca} R.,  {Caccianiga} A.,   {Severgnini}
  P.,  2012, \mn@doi [\mnras] {10.1111/j.1365-2966.2012.20445.x}, \href
  {http://adsabs.harvard.edu/abs/2012MNRAS.421.1803M} {421, 1803}

\bibitem[\protect\citeauthoryear{{Matt}, {Guainazzi}  \& {Maiolino}}{{Matt}
  et~al.}{2003}]{Matt03}
{Matt} G.,  {Guainazzi} M.,   {Maiolino} R.,  2003, \mn@doi [\mnras]
  {10.1046/j.1365-8711.2003.06539.x}, \href
  {http://adsabs.harvard.edu/abs/2003MNRAS.342..422M} {342, 422}

\bibitem[\protect\citeauthoryear{{Matt} et~al.,}{{Matt} et~al.}{2015}]{Matt15}
{Matt} G.,  et~al., 2015, \mn@doi [\mnras] {10.1093/mnras/stu2653}, \href
  {http://adsabs.harvard.edu/abs/2015MNRAS.447.3029M} {447, 3029}

\bibitem[\protect\citeauthoryear{{Miniutti}, {Fabian}, {Brandt}, {Gallo}  \&
  {Boller}}{{Miniutti} et~al.}{2009}]{Miniutti09_phl1092}
{Miniutti} G.,  {Fabian} A.~C.,  {Brandt} W.~N.,  {Gallo} L.~C.,   {Boller} T.,
   2009, \mn@doi [\mnras] {10.1111/j.1745-3933.2009.00669.x}, \href
  {http://adsabs.harvard.edu/abs/2009MNRAS.396L..85M} {396, L85}

\bibitem[\protect\citeauthoryear{{Miniutti}, {Saxton},
  {Rodr{\'{\i}}guez-Pascual}, {Read}, {Esquej}, {Colless}, {Dobbie}  \&
  {Spolaor}}{{Miniutti} et~al.}{2013}]{Miniutti13}
{Miniutti} G.,  {Saxton} R.~D.,  {Rodr{\'{\i}}guez-Pascual} P.~M.,  {Read}
  A.~M.,  {Esquej} P.,  {Colless} M.,  {Dobbie} P.,   {Spolaor} M.,  2013,
  \mn@doi [\mnras] {10.1093/mnras/stt850}, \href
  {http://adsabs.harvard.edu/abs/2013MNRAS.433.1764M} {433, 1764}

\bibitem[\protect\citeauthoryear{{Miniutti} et~al.,}{{Miniutti}
  et~al.}{2014}]{Miniutti14}
{Miniutti} G.,  et~al., 2014, \mn@doi [\mnras] {10.1093/mnras/stt2005}, \href
  {http://adsabs.harvard.edu/abs/2014MNRAS.437.1776M} {437, 1776}

\bibitem[\protect\citeauthoryear{{Nicastro}}{{Nicastro}}{2000}]{Nicastro00}
{Nicastro} F.,  2000, \mn@doi [\apjl] {10.1086/312491}, \href
  {http://adsabs.harvard.edu/abs/2000ApJ...530L..65N} {530, L65}

\bibitem[\protect\citeauthoryear{{Osterbrock} \& {Ferland}}{{Osterbrock} \&
  {Ferland}}{2006}]{osterbrock06}
{Osterbrock} D.~E.,  {Ferland} G.~J.,  2006, {Astrophysics of gaseous nebulae
  and active galactic nuclei}

\bibitem[\protect\citeauthoryear{{Parker} et~al.,}{{Parker}
  et~al.}{2014a}]{Parker14_mrk335}
{Parker} M.~L.,  et~al., 2014a, \mn@doi [\mnras] {10.1093/mnras/stu1246}, \href
  {http://adsabs.harvard.edu/abs/2014MNRAS.443.1723P} {443, 1723}

\bibitem[\protect\citeauthoryear{{Parker}, {Schartel}, {Komossa}, {Grupe},
  {Santos-Lle{\'o}}, {Fabian}  \& {Mathur}}{{Parker}
  et~al.}{2014b}]{Parker14_mrk1048}
{Parker} M.~L.,  {Schartel} N.,  {Komossa} S.,  {Grupe} D.,  {Santos-Lle{\'o}}
  M.,  {Fabian} A.~C.,   {Mathur} S.,  2014b, \mn@doi [\mnras]
  {10.1093/mnras/stu1818}, \href
  {http://adsabs.harvard.edu/abs/2014MNRAS.445.1039P} {445, 1039}

\bibitem[\protect\citeauthoryear{{Penston} \& {Perez}}{{Penston} \&
  {Perez}}{1984}]{Penston84}
{Penston} M.~V.,  {Perez} E.,  1984, \mnras, \href
  {http://adsabs.harvard.edu/abs/1984MNRAS.211P..33P} {211, 33P}

\bibitem[\protect\citeauthoryear{{Peterson} et~al.,}{{Peterson}
  et~al.}{1994}]{peterson94}
{Peterson} B.~M.,  et~al., 1994, \mn@doi [\apj] {10.1086/174009}, \href
  {http://adsabs.harvard.edu/abs/1994ApJ...425..622P} {425, 622}

\bibitem[\protect\citeauthoryear{{Pozo Nu{\~n}ez} et~al.,}{{Pozo Nu{\~n}ez}
  et~al.}{2015}]{Nunoz15}
{Pozo Nu{\~n}ez} F.,  et~al., 2015, \mn@doi [\aap]
  {10.1051/0004-6361/201525910}, \href
  {http://adsabs.harvard.edu/abs/2015A%26A...576A..73P} {576, A73}

\bibitem[\protect\citeauthoryear{{Predehl} \& {Schmitt}}{{Predehl} \&
  {Schmitt}}{1995}]{Predehl95}
{Predehl} P.,  {Schmitt} J.~H.~M.~M.,  1995, \aap, \href
  {http://adsabs.harvard.edu/abs/1995A%26A...293..889P} {293}

\bibitem[\protect\citeauthoryear{{Rees}}{{Rees}}{1988}]{Rees88}
{Rees} M.~J.,  1988, \mn@doi [\nat] {10.1038/333523a0}, \href
  {http://adsabs.harvard.edu/abs/1988Natur.333..523R} {333, 523}

\bibitem[\protect\citeauthoryear{{Rees}}{{Rees}}{1990}]{Rees90}
{Rees} M.~J.,  1990, \mn@doi [Science] {10.1126/science.247.4944.817}, \href
  {http://adsabs.harvard.edu/abs/1990Sci...247..817R} {247, 817}

\bibitem[\protect\citeauthoryear{{Reeves}, {Done}, {Pounds}, {Terashima},
  {Hayashida}, {Anabuki}, {Uchino}  \& {Turner}}{{Reeves}
  et~al.}{2008}]{Reeves08}
{Reeves} J.,  {Done} C.,  {Pounds} K.,  {Terashima} Y.,  {Hayashida} K.,
  {Anabuki} N.,  {Uchino} M.,   {Turner} M.,  2008, \mn@doi [\mnras]
  {10.1111/j.1745-3933.2008.00443.x}, \href
  {http://adsabs.harvard.edu/abs/2008MNRAS.385L.108R} {385, L108}

\bibitem[\protect\citeauthoryear{{Reimers}, {Koehler}  \& {Wisotzki}}{{Reimers}
  et~al.}{1996}]{Reimers96}
{Reimers} D.,  {Koehler} T.,   {Wisotzki} L.,  1996, \aaps, \href
  {http://adsabs.harvard.edu/abs/1996A%26AS..115..235R} {115, 235}

\bibitem[\protect\citeauthoryear{{Ricci} et~al.,}{{Ricci}
  et~al.}{2016}]{Ricci16}
{Ricci} C.,  et~al., 2016, \mn@doi [\apj] {10.3847/0004-637X/820/1/5}, \href
  {http://adsabs.harvard.edu/abs/2016ApJ...820....5R} {820, 5}

\bibitem[\protect\citeauthoryear{{Risaliti}, {Elvis}, {Fabbiano}, {Baldi}  \&
  {Zezas}}{{Risaliti} et~al.}{2005}]{Risaliti05}
{Risaliti} G.,  {Elvis} M.,  {Fabbiano} G.,  {Baldi} A.,   {Zezas} A.,  2005,
  \mn@doi [\apjl] {10.1086/430252}, \href
  {http://adsabs.harvard.edu/abs/2005ApJ...623L..93R} {623, L93}

\bibitem[\protect\citeauthoryear{{Ruan} et~al.,}{{Ruan} et~al.}{2015}]{Ruan15}
{Ruan} J.~J.,  et~al., 2015, preprint, \href
  {http://adsabs.harvard.edu/abs/2015arXiv150903634R} {} (\mn@eprint {arXiv}
  {1509.03634})

\bibitem[\protect\citeauthoryear{{Runco} et~al.,}{{Runco}
  et~al.}{2016}]{Runco16}
{Runco} J.~N.,  et~al., 2016, \mn@doi [\apj] {10.3847/0004-637X/821/1/33},
  \href {http://adsabs.harvard.edu/abs/2016ApJ...821...33R} {821, 33}

\bibitem[\protect\citeauthoryear{{Runnoe} et~al.,}{{Runnoe}
  et~al.}{2016}]{Runnoe16}
{Runnoe} J.~C.,  et~al., 2016, \mn@doi [\mnras] {10.1093/mnras/stv2385}, \href
  {http://adsabs.harvard.edu/abs/2016MNRAS.455.1691R} {455, 1691}

\bibitem[\protect\citeauthoryear{{Sakata} et~al.,}{{Sakata}
  et~al.}{2010}]{Sakata10}
{Sakata} Y.,  et~al., 2010, \mn@doi [\apj] {10.1088/0004-637X/711/1/461}, \href
  {http://adsabs.harvard.edu/abs/2010ApJ...711..461S} {711, 461}

\bibitem[\protect\citeauthoryear{{Saxton} et~al.,}{{Saxton}
  et~al.}{2014}]{Saxton14}
{Saxton} R.~D.,  et~al., 2014, \mn@doi [\aap] {10.1051/0004-6361/201424347},
  \href {http://adsabs.harvard.edu/abs/2014A%26A...572A...1S} {572, A1}

\bibitem[\protect\citeauthoryear{{Schartel}, {Rodr{\'{\i}}guez-Pascual},
  {Santos-Lle{\'o}}, {Ballo}, {Clavel}, {Guainazzi}, {Jim{\'e}nez-Bail{\'o}n}
  \& {Piconcelli}}{{Schartel} et~al.}{2007}]{Schartel07}
{Schartel} N.,  {Rodr{\'{\i}}guez-Pascual} P.~M.,  {Santos-Lle{\'o}} M.,
  {Ballo} L.,  {Clavel} J.,  {Guainazzi} M.,  {Jim{\'e}nez-Bail{\'o}n} E.,
  {Piconcelli} E.,  2007, \mn@doi [\aap] {10.1051/0004-6361:20077812}, \href
  {http://adsabs.harvard.edu/abs/2007A%26A...474..431S} {474, 431}

\bibitem[\protect\citeauthoryear{{Schartel}, {Rodr{\'{\i}}guez-Pascual},
  {Santos-Lle{\'o}}, {Jim{\'e}nez-Bail{\'o}n}, {Ballo}  \&
  {Piconcelli}}{{Schartel} et~al.}{2010}]{Schartel10}
{Schartel} N.,  {Rodr{\'{\i}}guez-Pascual} P.~M.,  {Santos-Lle{\'o}} M.,
  {Jim{\'e}nez-Bail{\'o}n} E.,  {Ballo} L.,   {Piconcelli} E.,  2010, \mn@doi
  [\aap] {10.1051/0004-6361/200912389}, \href
  {http://adsabs.harvard.edu/abs/2010A%26A...512A..75S} {512, A75}

\bibitem[\protect\citeauthoryear{{Schlafly} \& {Finkbeiner}}{{Schlafly} \&
  {Finkbeiner}}{2011}]{schlafly11}
{Schlafly} E.~F.,  {Finkbeiner} D.~P.,  2011, \mn@doi [\apj]
  {10.1088/0004-637X/737/2/103}, \href
  {http://adsabs.harvard.edu/abs/2011ApJ...737..103S} {737, 103}

\bibitem[\protect\citeauthoryear{{Schlegel}, {Finkbeiner}  \&
  {Davis}}{{Schlegel} et~al.}{1998}]{schlegel98}
{Schlegel} D.~J.,  {Finkbeiner} D.~P.,   {Davis} M.,  1998, \mn@doi [\apj]
  {10.1086/305772}, \href {http://adsabs.harvard.edu/abs/1998ApJ...500..525S}
  {500, 525}

\bibitem[\protect\citeauthoryear{{Shappee} et~al.,}{{Shappee}
  et~al.}{2014}]{Shappee14}
{Shappee} B.~J.,  et~al., 2014, \mn@doi [\apj] {10.1088/0004-637X/788/1/48},
  \href {http://adsabs.harvard.edu/abs/2014ApJ...788...48S} {788, 48}

\bibitem[\protect\citeauthoryear{{Storchi-Bergmann}, {Baldwin}  \&
  {Wilson}}{{Storchi-Bergmann} et~al.}{1993}]{Storchi-Bergmann93}
{Storchi-Bergmann} T.,  {Baldwin} J.~A.,   {Wilson} A.~S.,  1993, \mn@doi
  [\apjl] {10.1086/186867}, \href
  {http://adsabs.harvard.edu/abs/1993ApJ...410L..11S} {410, L11}

\bibitem[\protect\citeauthoryear{{Str{\"u}der} et~al.,}{{Str{\"u}der}
  et~al.}{2001}]{Struder01}
{Str{\"u}der} L.,  et~al., 2001, \mn@doi [\aap] {10.1051/0004-6361:20000066},
  \href {http://adsabs.harvard.edu/abs/2001A%26A...365L..18S} {365, L18}

\bibitem[\protect\citeauthoryear{{Titarchuk}}{{Titarchuk}}{1994}]{Titarchuk94}
{Titarchuk} L.,  1994, \mn@doi [\apj] {10.1086/174760}, \href
  {http://adsabs.harvard.edu/abs/1994ApJ...434..570T} {434, 570}

\bibitem[\protect\citeauthoryear{{Tohline} \& {Osterbrock}}{{Tohline} \&
  {Osterbrock}}{1976}]{Tohline76}
{Tohline} J.~E.,  {Osterbrock} D.~E.,  1976, \mn@doi [\apjl] {10.1086/182317},
  \href {http://adsabs.harvard.edu/abs/1976ApJ...210L.117T} {210, L117}

\bibitem[\protect\citeauthoryear{{Tombesi}, {Cappi}, {Reeves}, {Palumbo},
  {Yaqoob}, {Braito}  \& {Dadina}}{{Tombesi} et~al.}{2010}]{Tombesi10}
{Tombesi} F.,  {Cappi} M.,  {Reeves} J.~N.,  {Palumbo} G.~G.~C.,  {Yaqoob} T.,
  {Braito} V.,   {Dadina} M.,  2010, \mn@doi [\aap]
  {10.1051/0004-6361/200913440}, \href
  {http://adsabs.harvard.edu/abs/2010A%26A...521A..57T} {521, A57}

\bibitem[\protect\citeauthoryear{{Tran}, {Osterbrock}  \& {Martel}}{{Tran}
  et~al.}{1992}]{Tran92}
{Tran} H.~D.,  {Osterbrock} D.~E.,   {Martel} A.,  1992, \mn@doi [\aj]
  {10.1086/116382}, \href {http://adsabs.harvard.edu/abs/1992AJ....104.2072T}
  {104, 2072}

\bibitem[\protect\citeauthoryear{{Turner} et~al.,}{{Turner}
  et~al.}{2001}]{Turner01_MOS}
{Turner} M.~J.~L.,  et~al., 2001, \mn@doi [\aap] {10.1051/0004-6361:20000087},
  \href {http://adsabs.harvard.edu/abs/2001A%26A...365L..27T} {365, L27}

\bibitem[\protect\citeauthoryear{{Vasudevan}, {Mushotzky}, {Winter}  \&
  {Fabian}}{{Vasudevan} et~al.}{2009}]{Vasudevan09}
{Vasudevan} R.~V.,  {Mushotzky} R.~F.,  {Winter} L.~M.,   {Fabian} A.~C.,
  2009, \mn@doi [\mnras] {10.1111/j.1365-2966.2009.15371.x}, \href
  {http://adsabs.harvard.edu/abs/2009MNRAS.399.1553V} {399, 1553}

\bibitem[\protect\citeauthoryear{{Vestergaard} \& {Peterson}}{{Vestergaard} \&
  {Peterson}}{2006}]{Vestergaard06}
{Vestergaard} M.,  {Peterson} B.~M.,  2006, \mn@doi [\apj] {10.1086/500572},
  \href {http://adsabs.harvard.edu/abs/2006ApJ...641..689V} {641, 689}

\bibitem[\protect\citeauthoryear{{Voges} et~al.,}{{Voges}
  et~al.}{2000}]{Voges00}
{Voges} W.,  et~al., 2000, \iaucirc, \href
  {http://adsabs.harvard.edu/abs/2000IAUC.7432....3V} {7432, 3}

\bibitem[\protect\citeauthoryear{{Wang}, {Zhou}, {Wang}, {Lu}  \& {Xu}}{{Wang}
  et~al.}{2011}]{Wang11}
{Wang} T.-G.,  {Zhou} H.-Y.,  {Wang} L.-F.,  {Lu} H.-L.,   {Xu} D.,  2011,
  \mn@doi [\apj] {10.1088/0004-637X/740/2/85}, \href
  {http://adsabs.harvard.edu/abs/2011ApJ...740...85W} {740, 85}

\bibitem[\protect\citeauthoryear{{Willingale}, {Starling}, {Beardmore},
  {Tanvir}  \& {O'Brien}}{{Willingale} et~al.}{2013}]{Willingale13}
{Willingale} R.,  {Starling} R.~L.~C.,  {Beardmore} A.~P.,  {Tanvir} N.~R.,
  {O'Brien} P.~T.,  2013, \mn@doi [\mnras] {10.1093/mnras/stt175}, \href
  {http://adsabs.harvard.edu/abs/2013MNRAS.431..394W} {431, 394}

\bibitem[\protect\citeauthoryear{{Wilms}, {Allen}  \& {McCray}}{{Wilms}
  et~al.}{2000}]{Wilms00}
{Wilms} J.,  {Allen} A.,   {McCray} R.,  2000, \mn@doi [\apj] {10.1086/317016},
  \href {http://adsabs.harvard.edu/abs/2000ApJ...542..914W} {542, 914}

\bibitem[\protect\citeauthoryear{{Winkler}, {Glass}, {van Wyk}, {Marang},
  {Jones}, {Buckley}  \& {Sekiguchi}}{{Winkler} et~al.}{1992}]{Winkler92}
{Winkler} H.,  {Glass} I.~S.,  {van Wyk} F.,  {Marang} F.,  {Jones} J.~H.~S.,
  {Buckley} D.~A.~H.,   {Sekiguchi} K.,  1992, \mn@doi [\mnras]
  {10.1093/mnras/257.4.659}, \href
  {http://adsabs.harvard.edu/abs/1992MNRAS.257..659W} {257, 659}

\bibitem[\protect\citeauthoryear{{Zoghbi} et~al.,}{{Zoghbi}
  et~al.}{2015}]{Zoghbi15}
{Zoghbi} A.,  et~al., 2015, \mn@doi [\apjl] {10.1088/2041-8205/799/2/L24},
  \href {http://adsabs.harvard.edu/abs/2015ApJ...799L..24Z} {799, L24}

\makeatother
\end{thebibliography}

\end{document}